\documentclass[utf8]{FrontiersinHarvard} 

\usepackage{url,hyperref,lineno,microtype}
\usepackage[onehalfspacing]{setspace}

\newcommand\HI{H{\sc i}}

\newcommand{\bperp}{$B_{\textsc{POS}}$} 
\newcommand{\blos}{$B_{\textsc{LOS}}$} 

\newcommand{\ch}{}

\def\firstAuthorLast{Tahani} 
\def\Authors{Mehrnoosh Tahani\,$^{1, 2,*}$}

\def\Address{$^{1}$Dominion Radio Astrophysical Observatory, Herzberg Astronomy and Astrophysics Research Centre, National Research Council Canada, P. O. Box 248, Penticton, BC V2A 6J9 Canada }

\def\Address{$^{1}$Banting and KIPAC Fellow: Kavli Institute for Particle Astrophysics \& Cosmology (KIPAC), Stanford University, Stanford, CA 94305, USA\\
$^{2}$Covington Fellow: Dominion Radio Astrophysical Observatory, Herzberg Astronomy and Astrophysics Research Centre, National Research Council Canada, P. O. Box 248, Penticton, BC V2A 6J9 Canada}

\def\corrAuthor{Mehrnoosh Tahani}

\def\corrEmail{mtahani@stanford.edu}

\begin{document}
\onecolumn
\firstpage{1}

\title[3D fields]{Three-dimensional magnetic fields of molecular clouds}

\author[\firstAuthorLast ]{\Authors} 
\address{} 
\correspondance{} 

\extraAuth{}

\maketitle

\begin{abstract}
To investigate the role of magnetic fields in the evolution of the interstellar medium, formation and evolution of molecular clouds, and ultimately the formation of stars, their three-dimensional (3D) magnetic fields must be probed. Observing only one component of magnetic fields (along the line of sight or parallel to the plane of the sky)  is insufficient to identify these 3D vectors. In recent years, novel techniques for probing each of these two components and integrating them with additional data (from observations or models), such as Galactic magnetic fields or magnetic field inclination angles, have been developed, in order to infer 3D magnetic fields. We review and discuss these advancements, their applications, and their future direction.
\end{abstract}

\section{Introduction}

\ch{The Gaia mission~\citep{GaiaMission2016}, particularly with its stellar parallax distances~\citep[][]{Lurietal2018} and radial velocities~\citep[][]{Soubiranetal2018}, has enabled significant advances in various areas of astrophysics, 
ranging from the Galaxy structure~\citep[e.g.,][]{KounkelCovey2019} and evolution~\citep[e.g.,][]{Poggioetal2020, RuizLaraetal2020} to binary systems~\citep[][]{Wyrzykowskietal2020}. 
Thanks to Gaia, the three-dimensional (3D) density field of the Galaxy, especially of nearby molecular clouds~\citep{Grossschedletal2018, Zuckeretal2021, Rezaeietal2020, RezaeiKainulainen2022} and the solar neighborhood~\citep[e.g.,][]{Zuckeretal2022}, can now be mapped, enabling us to study the interstellar medium (ISM) evolution~\citep[e.g.,][]{Bialy2021, Kounkeletal2022}. 
However, studies of the ISM evolution are incomplete without observing 3D magnetic fields, as the two are interdependent~\citep[e.g.,][]{Tahanietal2022P, Kounkeletal2022}.}

\ch{Magnetic fields influence the star-formation process, from the evolution of diffuse ISM~\citep{Haverkorn2015} and formation of molecular clouds~\citep[e.g.,][]{Iwasakietal2019} to formation of sub-structures and stars~\citep[e.g.,][and references therein]{Pattleetal2022PP7}. However, their role remains undetermined~\citep[e.g,][]{HennebelleInutsuka2019, KrumholzFederrath2019}. 
Magnetic fields can stabilize  the clouds against gravity~\citep[e.g.,][]{FiegePudritzI2000, FiegePudritzII2000}, allow for the formation of denser structures and stars~\citep[e.g.,][]{Inoueetal2018}, reduce the star-formation rate~\citep[see][and references therein]{HennebelleInutsuka2019}, or regulate gas flow~\citep[e.g.,][]{SeifriedWalch2015}.  }

\ch{Magnetic field orientation relative to density structures may indicate their role in the ISM evolution~\citep[e.g.,][]{SolerHennebelle2017}.
Observations of plane-of-sky magnetic fields~\citep[\bperp ; e.g.,][]{PlanckXXXV, Soler2019} show that they tend to be perpendicular to high-column-density structures ($N_H > 10^{21.7}$) and parallel to low-column-density ones.  The relation between the transition from parallel to perpendicular alignment and gravitational collapse or Alfv\'en Mach number ($\mathcal{M}_A$) is being studied~\citep[e.g.,][]{Chenetal2016, SolerHennebelle2017, Soleretal2017, Pattleetal2021}. 3D magnetic field measurements are necessary to understand this alignment~\citep[][particularly since field lines may be inclined along the line of sight]{Girichidis2021}. }

\ch{The magnetic field inclination angle with respect to the plane of the sky ($\gamma$) also complicates inferring $\mathcal{M}_A$ and mass-to-flux ratio ($\mu_{\phi}$), two key quantities in examining the role of magnetic fields in star formation. $\mathcal{M}_A$ and $\mu_{\phi}$ quantify the cloud's magnetic energy relative to its kinetic/turbulent and gravitational energies, respectively~\citep[see][and references therein]{Pattleetal2021}. Without estimating the 3D fields, a sub-Alfv\'enic cloud ($\mathcal{M}_A <1$; with highly inclined ordered field lines) may be misinterpreted as a 
super-Alfv\'enic cloud~\citep[$\mathcal{M}_A >1$ with tangled field lines dominated by the flow;][]{FalcetaGoncalvesetal2008}. Additionally, using field strengths based on a single component instead of 3D vectors may lead to incorrect estimates of $\mu_{\phi}$ and the relationship between the cloud's magnetic field and gravitational energies~\citep[][]{Crutcheretal2010, MouschoviasTassis2010, Clemensetal2016, Pillaietal2016}. }

Moreover, 3D magnetic field observations allow comparison of their morphologies to cloud-formation model predictions, enabling us to investigate ISM evolution and molecular-cloud formation. While observed magnetic field morphologies are consistent with some cloud-formation models~\citep[e.g.,][]{InoueFukui2013, Inutsukaetal2015, Inutsuka2016IAU, Gomezetal2018, Inoueetal2018, Abeetal2021}, observing the 3D magnetic fields of a large number of clouds is required to study their formation scenario and determine how magnetic fields influence the evolution of these clouds into filaments, cores, and, eventually, stars. 

Despite the rise of recent techniques to observe interstellar magnetic fields~\citep[][]{Clarketal2014, GonzalezCasanovaLazarian2017, Tahanietal2018, LazarianYuen2018, Huetal2019}, probing the 3D fields remains exceedingly challenging. 
To infer the 3D fields, observations  of both  line-of-sight magnetic fields (\blos ) and \bperp\ are required. 
\ch{Common techniques for observing the interstellar magnetic fields~\citep[see][and references therein for details]{Pattleetal2022PP7} include Zeeman splitting~\citep{CrutcherKemball2019}, Faraday rotation~\citep{Brown2008}, dust emission polarization~\citep{Draine2003}, starlight (dust extinction) polarization~\citep{Voshchinnikov2012}, and synchrotron emission~\citep{Beck2015}.   This mini-review focuses on \emph{molecular clouds\footnote{A number of recent studies have examined the 3D magnetic fields of the diffuse ISM~\citep[e.g.,][]{Ferriere2016, VanEcketal2017, Alvesetal2018, Panopoulouetal2019, ClarkHensley2019, Hensleyetal2019}.}} (a few to $\sim 100$\,pc).  
For molecular clouds,  }
Zeeman splitting and the Faraday-based technique of  \citet{Tahanietal2018} provide \blos , while dust emission and starlight polarization provide \bperp . We present techniques for probing the 3D magnetic fields of molecular clouds in Section~\ref{sec:methods} and discuss their applications and future directions in Section~\ref{sec:discussion}. 

\section{3D Magnetic fields}
\label{sec:methods}
Several methods \citep[e.g.,][]{Chenetal2019, Huetal2021, Huetal2021PPV, Tahanietal2019,  Tahanietal2022O, Tahanietal2022P, HuLazarian2022} have examined the 3D magnetic fields  of molecular clouds. \citet{Chenetal2019} and \citet{HuLazarian2022} use  \bperp\ (dust polarization) observations and their polarization fraction ($p$) \ch{to recover the mean inclination of the ordered\footnote{ordered: ignoring the random component due to turbulence or smaller-scale variations} magnetic fields of molecular clouds}, whereas \citet[][scales of a few to $\sim 100$\,pc]{Tahanietal2022O, Tahanietal2022P} incorporate \blos\  and \bperp\ observations along with Galactic magnetic field (GMF) models. 

Numerous observatories, including the James Clark Maxwell Telescope~\citep[JCMT; e.g.,][]{Eswaraiahetal2021, Ngocetal2021, Hwangetal2021, Kwonetal2022}, Planck Space Observatory~\citep[e.g.,][]{PlanckXXXV, Alinaetal2019}, Atacama Large Millimeter/sub-millimeter Array~\citep[ALMA; e.g.,][]{Pattleetal2021, Cortesetal2021}, Sub-Millimeter Array~\citep[SMA; e.g.,][]{Zhangetal2014}, {\ch{and Stratospheric Observatory for Infrared Astronomy~\citep[SOFIA; e.g.,][]{Chussetal2019}}} have observed \bperp\ of numerous star-forming regions.  However, the number of \blos\ observations of molecular clouds are still limited. Although Zeeman splitting is a powerful technique for probing  \blos\ and the most accurate method for determining field strengths, it requires lengthy observing runs, making it challenging to observe. The observing technique of~\citet{Tahanietal2018} can be used to map \blos\ of numerous molecular clouds. 

\subsection{Line-of-sight magnetic fields}

\citet{Tahanietal2018} developed a new technique to probe \blos\ associated with molecular clouds, using Faraday rotation. We provide a brief summary of the technique in this section. 

\subsubsection{Faraday rotation}

Due to the lower abundance  of electrons in molecular clouds \ch{(compared to ionized regions)}, it was previously believed that Faraday rotation\footnote{A number of review articles discuss Faraday rotation and its observations~\citep[e.g.,][]{Brown2008, Noutsos2012, Han2017}.}  could not be used to investigate the magnetic fields of molecular clouds. \citet{Tahanietal2018} developed a technique to successfully determine  \blos\ of molecular clouds using Faraday rotation measures (RM), while previous attempts~\citep[][]{Reichetal2002, WollebenReich2004} were unable to provide a map of \blos\ observations across the cloud.  

\subsubsection{Methodology and results}
\label{sec:blos}
In this technique~\citep{Tahanietal2018}, the non-cloud (background and foreground; Galactic) contribution to the RM \ch{(RM$_{\rm{ref}}$)} is subtracted from the observed RM of extra-galactic point sources (radio galaxies or quasars) using an on-off approach. Numerous catalogs~\citep[e.g.,][]{Tayloretal2009} provide observed RM point sources. Following the determination of the the cloud's RMs, the electron column densities associated with each RM point are calculated using a chemical evolution code and extinction maps. Any chemical evolution code~\citep[e.g., one used by][]{Gibsonetal2009} and extinction map~\citep[e.g.,][]{Kainulainenetal2009}, or even Hydrogen column density map~\citep[][]{Lombardi2014, Zarietal2016}, can be utilized. \ch{To find electron column densities}, the cloud is divided into sub-layers \ch{aligned along the line of sight} using extinction values and the chemical code. The electron column density in each sub-layer is obtained separately.  Calculating the average \blos\ along the line of sight is made possible by adding the electron column density contributions of these sub-layers.

\begin{figure}
\centering
\includegraphics[scale=0.36, trim={0.cm 0cm 0.5cm 0.cm},clip]{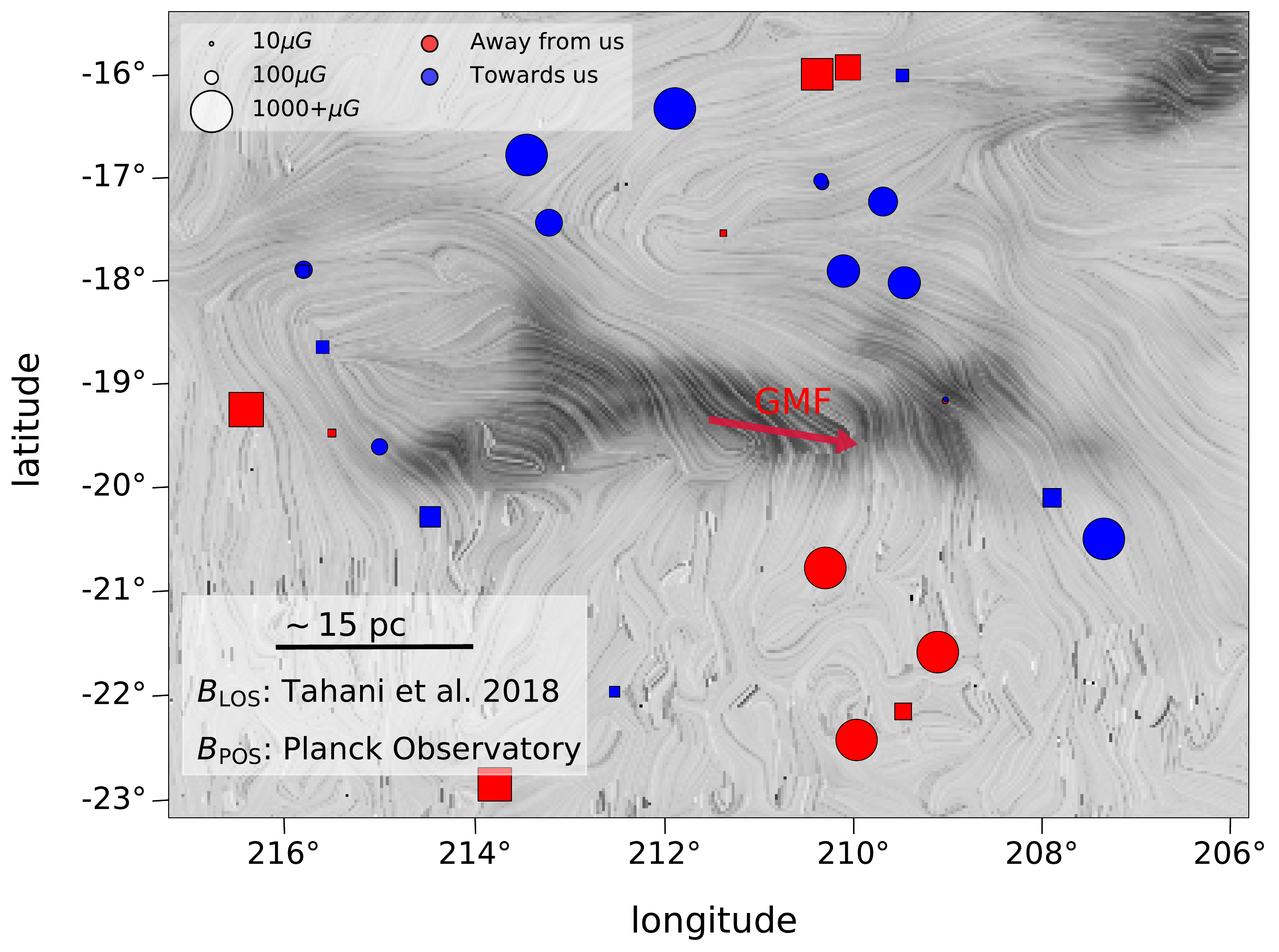}
\includegraphics[scale=0.52, trim={0cm 0cm 0cm 0cm},clip]{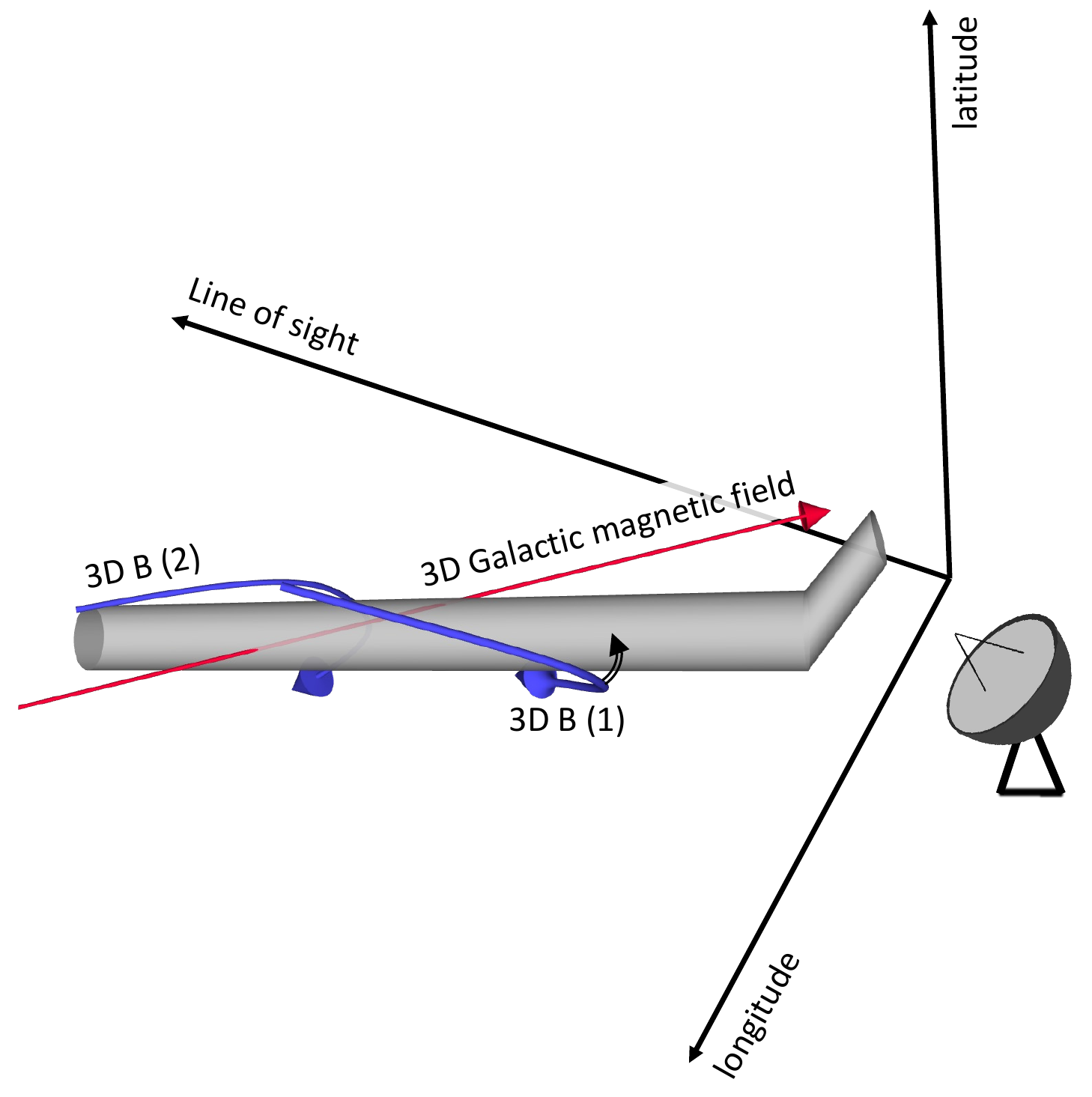}\\
\caption{3D magnetic field of the Orion A cloud. \textbf{Left panel:} The grayscale image illustrates the hydrogen column density map of Orion A~\citep{Lombardi2014}. The circle and square markers represent \blos , with the square indicating non-detection points (with high uncertainties that may cause a change in \blos\ direction) and blue (red) representing pointing toward (away from) us. The drapery lines represent the \bperp\ observed by the Planck Space Observatory. The red vector depicts the modeled Galactic Magnetic field projected onto the plane of the sky. The same \blos\ reversal throughout the cloud was previously detected using Zeeman measurements~\citep[][see their Figure ~15]{Heiles1997}. We note that in Zeeman measurements, the negative sign indicates magnetic field directed toward us, while in RM studies, it indicates magnetic field directed away from us.  \textbf{Right panel:} From our vantage point, the inferred 3D ordered magnetic field of Orion A is semi-convex. The red vector, bent gray cylinder, and blue vectors represent the modeled GMF, cloud, and 3D magnetic field of the cloud, respectively. Without identifying the inclination angle of the cloud, rotations of up to 50$^{\circ}$ along the black arrow may be possible, resulting in both B(1) and B(2)~\citep[see Section 2 of][]{Tahanietal2022O}.} 
\label{fig:Blos3D}
\end{figure}

\citet{Tahanietal2018} mapped \blos\ of the Orion A, Orion B, California, and Perseus molecular clouds and found that their results were consistent with existing molecular 
Zeeman measurements. \ch{They found that the \blos\ direction of the Orion A (see left panel of Figure~\ref{fig:Blos3D}) and California clouds  reverses from one side to the other (along the short axis of the cloud). Their Perseus results suggested a weak indication of this reversal.} The \blos\ reversal across Orion A was previously observed via Zeeman splitting~\citep{Heiles1997}, in the same
directions as~\citet{Tahanietal2018}.

\ch{Identifying a) direction and b) strength are the two components of \blos\ determination in this technique. The direction uncertainty arises from uncertainties in a) observed RM values and b) RM$_{\rm{ref}}$. The strength uncertainty arises from assumptions of a) constant \blos\ along the line of sight, b) symmetry of the cloud along the line of sight, c) parameters taken to estimate electron densities (cloud's initial temperature and density and Ultra-Violet and cosmic ionization rates), and d) extinction maps. 
}

\subsection{Plane of sky magnetic fields}

\ch{Dust emission polarization has been successfully applied to molecular clouds~\citep[e.g.,][]{PlanckXXXV, PattleFissel2019}.  The technique is based on the alignment of the long axis of amorphous dust grains~\citep[e.g.,][]{Draine2009} perpendicular to magnetic fields, resulting in linear polarization and explained by radiative torque alignment~\citep[RAT;][]{Draineetal1997, Lazarian2007, LazarianHoang2007, Anderssonetal2015, HoangLazarian2016}. The Davis-Chandrasekhar-Fermi technique~\citep[DCF;][]{DavisGreenstein1951, ChandrasekharFermi1953} or its subsequent modified versions~\citep[e.g.,][]{Ostrikeretal2001, Houdeetal2009, SkalidisTassis2021, Skalidisetal2021a} are utilized to estimate \bperp\ strengths~\citep[see][for more information and the technique's limitations]{PattleFissel2019, Pattleetal2022PP7}. 
}

\subsection{Reconstructing the mean 3D magnetic fields of molecular clouds}
\label{sec:3DReconstruct}

Using \blos\ observations, \citet{Tahanietal2019, Tahanietal2022O, Tahanietal2022P} studied the 3D magnetic field morphologies of the Orion A and Perseus molecular clouds. 
\ch{\citet{Tahanietal2019} constrained models of the ordered, cloud-scale magnetic field, using \bperp\ angles and \blos\ estimates, 
whereas \citet{Tahanietal2022O, Tahanietal2022P} inferred cloud-scale magnetic field vectors in 3D\footnote{approximate 3D morphology at scales of a few to 100\,pc (ignoring turbulence and smaller-scale variations)}, 
given a set of model assumptions.} We discuss these techniques in this section.

\subsubsection{Analytical models of the ordered magnetic field within clouds and comparison to synthetic observations}

\citet{Tahanietal2019} constructed models \ch{that could explain the observed \blos\ reversal discussed in Section~\ref{sec:blos}}, obtained synthetic observations from the models, and compared these synthetic observations with \blos\ (direction and strengths) and  \bperp\ (angle and strength; using Planck\footnote{\url{http://www.esa.int/Planck}}) estimates of Orion A. They concluded that an arc-shaped morphology (see right panel of Figure~\ref{fig:Blos3D}) is the most probable magnetic morphology for Orion A, based on Monte-Carlo analysis, chi-square probability values, and examination of a range of systematic biases between \blos\ and \bperp\ observations. In the arc-shaped morphology, field lines bend around the filamentary cloud in response to environmental interaction~\citep[first proposed by][]{Heiles1997}, enabling mass to flow along the field lines and accumulate on the cloud~\citep{Inoueetal2018}.

\subsubsection{Using Galactic magnetic field models to reconstruct the cloud-scale ordered magnetic field 3D vector}

\citet{Tahanietal2022O, Tahanietal2022P} \ch{reconstructed the cloud-scale, ordered magnetic field vectors} of the Orion A and Perseus clouds in 3D. Using  \blos\ and \bperp\ observations, along with large-scale  GMF models~\citep{JanssonFarrar2012, JanssonFarrar2012b}, they \ch{inferred} the \ch{approximate} orientation and direction\footnote{In this mini-review we distinguish between the terms \emph{direction} and \emph{orientation}. Knowing the direction reveals orientation, but not the other way around. For example, the direction of \blos\ indicates either away from us or toward us, whereas the orientation of \blos\ indicates only that the line is parallel to the line of sight without specifying its direction. Similarly for \bperp, direction refers to the complete 2D vector, while orientation refers only to the line without specifying the vector's endpoint.} of the 3D \ch{ordered} magnetic field of these clouds (including their \bperp\ \emph{direction}). Although the \bperp\ \emph{orientation} of numerous molecular clouds had been observed previously, their \bperp\ \emph{direction} remained undetermined even in the 3D study by \citet{Tahanietal2019}. 

Moreover, \ch{by estimating $\mathcal{M}_{A}$ values and/or comparing estimates of initial magnetic field vectors (using GMF models) with \bperp\ maps}, \citet{Tahanietal2022O, Tahanietal2022P}  suggest that the magnetic fields of the Orion A and Perseus clouds retain a memory of the Galactic magnetic fields. Although some studies~\citep[e.g.,][]{Stephensetal2011} have suggested that the magnetic fields of molecular clouds  are dissociated from larger Galactic scales, others~\citep[e.g.,][]{HanZhang2007} have concluded that they largely retain the large-scale Galactic magnetic fields.

\ch{We note that this technique relies on correctly identifying the ordered GMF vector at the cloud location.  This vector provides an approximation of the initial magnetic fields prior to the cloud's evolution (allowing us to ignore the GMF random component caused by cloud-scale turbulence). Since GMF models vary~\citep[][]{Jaffe2019}, this technique is applied to clouds in a region of the Galaxy (pointing anti-Galactic and nearby) where there is  less disagreement between the GMF models. For example, all models in Figure 2 from \citet{Jaffe2019}, except panel h~\citep{Fauvetetal2011}, generate similar ordered GMF vectors at the locations of the Orion A and Perseus clouds. Moreover, the limited number of \blos\ observations per cloud and the use of two tracers (dust emission and a Faraday-based technique) may increase the technique's uncertainties. Upcoming observations are required to advance these studies (see Section~\ref{sec:discussion}).  }

\subsection{Inclination angle: statistical studies of polarization fraction}
\label{sec:pfInclination}

The 3D morphologies identified by \citet{Tahanietal2022O, Tahanietal2022P} can be improved by inferring $\gamma$ at various points across the cloud and combining their method with studies that estimate $\gamma$ \citep[e.g.,][]{Chenetal2019, Sullivan2021, Huetal2021PPV, HuLazarianWang2022, HuLazarian2022}. 
In recent years, $\gamma$ has been inferred in molecular clouds~\citep[e.g.,][]{Sullivan2021} \ch{and diffuse ISM\footnote{where dust emission intensity per atomic hydrogen column density may also be used to infer $\gamma$~\citep{Hensleyetal2019}. }~\citep[e.g.,][]{Hensleyetal2019}, using the dependence of $p$ and polarization angle dispersion ($\mathcal{S}$) on $\gamma$~\citep[e.g.,][]{FalcetaGoncalvesetal2008, Hensleyetal2019}, under the assumption of homogeneous grain alignment efficiency. 
}

\citet{Kingetal2018} compared the $p$ and $\mathcal{S}$ values of the Vela C cloud with their 3D, ideal magnetohydrodynamics (MHD) colliding flow simulations. The simulations were performed using the ATHENA code \citep{Stoneetal2008} and included gravity. 
Statistical comparisons \ch{(using relative orientation of column density and magnetic fields, average $\gamma$, and $\mathcal{S}$)} between these simulations and observations explored the effect of $\gamma$ on $p$ and $\mathcal{S}$ and were made possible by the high resolution and sensitivity of the Balloon-borne Large Aperture Sub-millimeter Telescope for Polarimetry (BLASTPol) observations of the Vela C~\citep{Fisseletal2016} cloud. 
These comparisons indicated that the Vela C observations and its high polarization angle dispersion were consistent with simulations of magnetic  fields with high inclination angles. \ch{However, due to the degeneracy between disorder caused by turbulence and disorder caused by a large inclination angle} (the field disorder seen in the plane of the sky), they were unable to infer a $\gamma$ value for the Vela C cloud. 

\citet{Chenetal2019} extended the study of \citet{Kingetal2018} and determined $\gamma$ for the Vela C cloud, assuming a small total $\mathcal{S}$ (applicable only to sub-Alfv\'enic regions). Using a statistical examination of the $p$ values of the cloud and the maximum polarization fraction (associated with zero inclination), they calculated $\gamma$. They found an average $\gamma$ value of $\sim 60^{\circ}$ for the Vela C cloud, with an estimated accuracy of $\leq 10^{\circ} - 30^{\circ}$. Subsequently, \citet{Sullivan2021} analyzed the 3D magnetic field  properties of nearby molecular clouds\footnote{The Aquila Rift, Cepheus, Chameleon-Musca, Corona Australis, Lupus, Ophiuchus, Perseus, Taurus, and Vela C clouds} and estimated their cloud-averaged $\gamma$ values. This technique can be used to examine the relative alignment of magnetic field lines and the orientation of filamentary dense gas in 3D~\citep{Fisseletal2019}. 

\begin{figure}
\centering
\includegraphics[scale=0.55, trim={0.cm 0.5cm 0.5cm 0.5cm},clip]{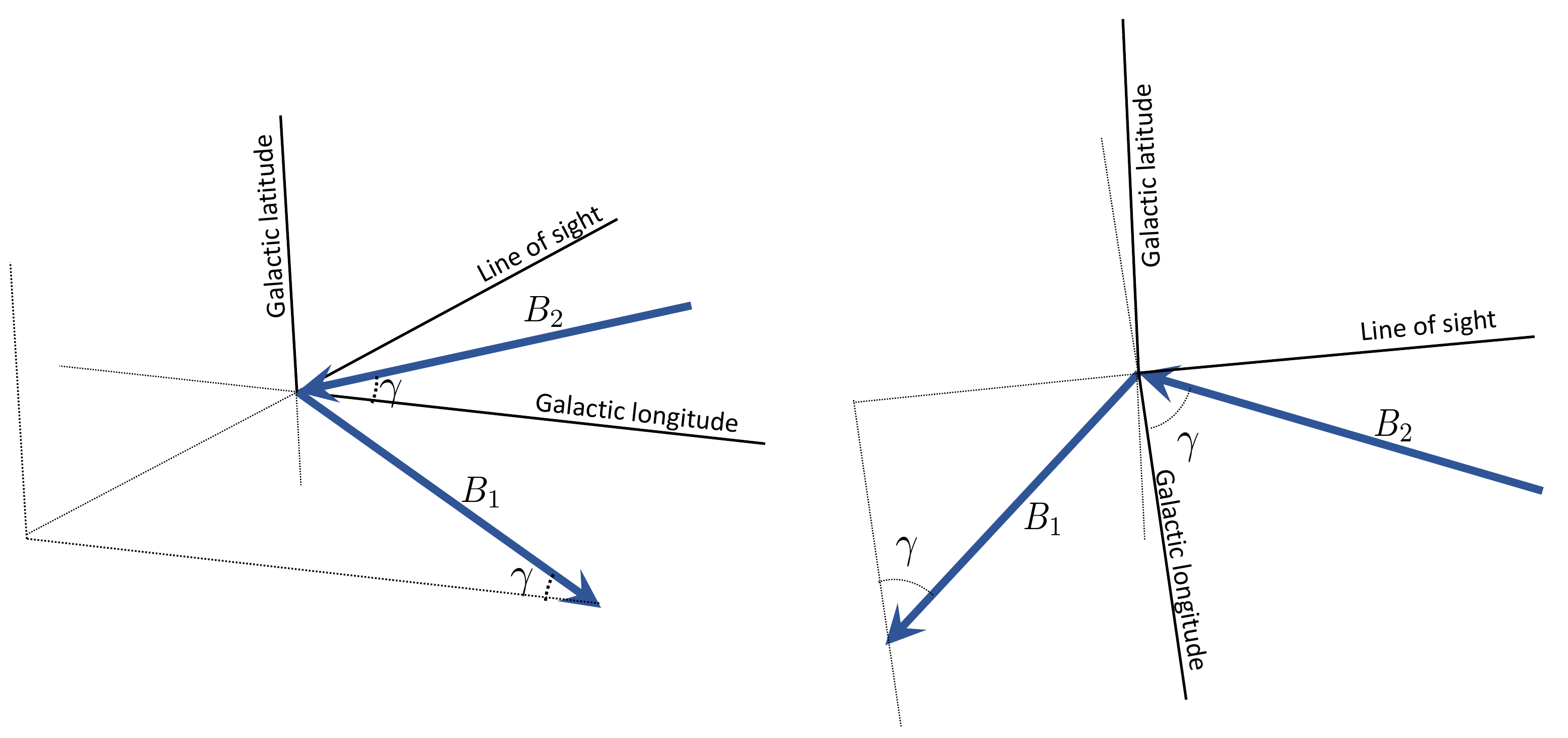}
\caption{\bperp\ direction required for 3D field determination. The 3D magnetic field vectors $B_1$ and $B_2$ have the same inclination angle ($\gamma$), run parallel to the Galactic longitude axis when projected onto the plane of the sky, and point toward us when projected along the line of sight. However, due to the difference in their \bperp\ directions, they are two distinct 3D vectors. 
Since the projections of these two vectors onto the plane of the sky are parallel to the longitude axis, their inclination angle with respect to the plane of the sky is the angle between the 3D vector and the longitude axis. The left and right panels display two different viewing angles.  \ch{Distinguishing between these two vectors is particularly important in studies of relative alignment of field lines and clouds, as a cloud aligned with $B_1$ may be approximately perpendicular to $B_2$ depending on the value of $\gamma$.}  } 
\label{fig:inclination}
\end{figure}

\ch{The technique's inherent uncertainty is dominated by the following assumptions: a) presence of a location within the cloud with zero $\gamma$, corresponding to the observed maximum $p$; b) homogeneous grain alignment efficiency across the cloud\footnote{\citet{Kingetal2019} suggest that the correlation between $\mathcal{S}$ and $\gamma$ used in the~\citet{Chenetal2019} technique is maintained, even in the absence of homogeneous grain alignment efficiency, assuming a power-law relation between grain alignment efficiency and local gas density.}; c) neglecting depolarization effects along the line of sight; d) assuming uni-directional magnetic fields along the line of sight; and e) ordered field line, which was addressed by \citet{HuLazarian2022}.} \citet{HuLazarian2022} augmented the technique of \citet{Chenetal2019} by incorporating  magnetic field fluctuations and dispersion (making the technique applicable to trans- and super-Alfv\'enic regions as well). They modified the equations of \citet{Chenetal2019} on the assumption that field fluctuations are perpendicular to the mean field. 
Additionally, we note that these studies still require both \blos\ and \bperp\ \emph{directions} to infer 3D vectors (see Figure~\ref{fig:inclination}).

\subsection{Other approaches}

\ch{While this mini-review focuses on the techniques discussed in Sections~\ref{sec:3DReconstruct} and \ref{sec:pfInclination} and their combination for recovering the 3D magnetic fields of \emph{molecular clouds}, 
we note that other more theory-based techniques can also be used in clouds~\citep[e.g.,][]{YanLazarian2005, TritsisTassis2018, Huetal2021PPV, Skalidisetal2021} or within its high density regions~\citep[i.e., clumps or cores][]{Houdeetal2000a, Kandorietal2017, Kandorietal2020_3Da, Kandorietal2020_3Db, Kandorietal2020_3Dc}. We briefly discuss these techniques here, excluding those applicable only to core scales~\citep[e.g.,][]{Kandorietal2017, Kandorietal2020_3Da, Kandorietal2020_3Db, Kandorietal2020_3Dc}. }

\subsubsection{Ion-to-neutral line-width}

\cite{Houdeetal2000a, Houdeetal2000b, Houdeetal2002, Houdeetal2004} proposed a method for measuring $\gamma$ based on the ion-to-neutral line-width ratios. Their observations showed that, in the presence of strong magnetic fields, the line-width of ions is narrower than that of coexisting neutrals. They suggest that when the field lines are perpendicular to the line of sight, the difference in line-widths should be the greatest, enabling them to infer $\gamma$.  Some studies found supporting \citep{LiHoude2008, Hezarehetal2010, Houde2011, Tangetal2018} or inconsistent \citep{Pinedaetal2021} observational evidence.

\subsubsection{Atomic alignment}

\ch{The atomic alignment (or ground state alignment) technique~\citep{YanLazarian2005, YanLazarian2006, YanLazarian2007, YanLazarian2012, Yanetal2019} relies on the alignment of the angular momentum of atoms in their ground state with the photons’ angular momentum from background anisotropic radiation, followed by their realignment with external magnetic fields.  
For best outcomes, absorption lines are used. Calculating the degree of alignment with magnetic field lines, \citet{YanLazarian2007} obtained the Stokes parameters of absorbed radiation and compared them with observations to infer $\gamma$ and the 3D field lines.  
This method is most applicable to  diffuse ISM~\citep{YanLazarian2012}, but may also be applied to molecular clouds and their envelopes. }

\subsubsection{Young stellar objects and position-position-velocity space techniques}

\ch{Based on the observable anisotropy of turbulence eddies in the presence of magnetic fields, \citet{Huetal2021} estimate magnetic fields using structure function analysis (SFA). They demonstrate that for sub-Alfv\'enic regions, the ratio of perpendicular to parallel\footnote{relative to the magnetic field} velocity fluctuations has a power-law relation with $\mathcal{M}_A$, enabling determination of 3D field strengths. 
\citet{Huetal2021PPV} extended the SFA analysis of \citet{Huetal2021} to infer 3D fields by incorporating Gaia observations of young stellar objects (for estimating 3D velocity fluctuations; assuming they inherit the velocity of their parent cloud). }

\subsection{Potential insights  from 3D field mapping}

\ch{This section briefly discusses the potential takeaways from the aforementioned 3D studies. 
Assuming a GMF model and given \blos\  and \bperp\ observations,  \citet{Tahanietal2022P, Tahanietal2022O} inferred the 3D ordered magnetic field vectors of two molecular clouds. Including $\gamma$ can enhance these studies. Inferring the 3D magnetic fields of numerous molecular clouds will enable us to compare them with models and numerical simulations to constrain cloud formation models~\citep[see][and references therein]{HennebelleInutsuka2019}, 3D structure and evolution of the ISM~\citep[e.g.,][]{Hacaretal2022}, 3D GMF models~\citep[e.g.,][]{Jaffe2019}, and the role of magnetic fields in cloud evolution~\citep[e.g.,][]{FiegePudritzI2000}.} 

For example, \citet{Tahanietal2022P} employed velocity information of the Perseus cloud along with GMF models to predict the cloud-averaged ordered line-of-sight and 3D magnetic field of this cloud based on the model of \citet{Inutsukaetal2015}\footnote{also see simulations by \citet{Inoueetal2018}} and found the predictions to be consistent with their inferred 3D field and \blos\ data. The cloud-formation model of \citet{Inutsukaetal2015} requires multiple compressions caused by expanding interstellar bubbles to form filamentary molecular clouds.  Using dynamics and bubble observations of the Orion A and Perseus clouds, \citet{Tahanietal2022O, Tahanietal2022P} proposed similar formation scenarios for their 3D fields: the field lines should have been initially bent on a large scale by recurrent supernovae shocks. This bending of field lines by bubbles has been detected in numerical simulations~\citep{KimOstriker2015} and large- and small-scale observations~\citep{Soleretal2018, Braccoetal2020, Arzoumanianetal2021}. Subsequently, interaction with a secondary bubble may have pushed the \HI\ gas surrounding the clouds, causing a sharp field line bending (arc-shaped field) associated with the molecular cloud. 

\ch{Velocity profile observations may also shed light on the formation process or 3D structure of clouds~\citep[e.g.,][]{TritsisTassis2018, Arzoumanianetal2018, Bonneetal2020}.} Position-position-velocity space studies of these clouds can improve the precision and accuracy of these 3D fields to explore their consistency with theoretical and numerical  models~\citep[e.g.,][]{Clarketal2014, Clarketal2015, GonzalezCasanovaLazarian2017, Clark2018, ClarkHensley2019, Huetal2019, Huetal2020, Huetal2021PPV, Huetal2021, HuLazarianWang2022}.

\section{Discussion}
\label{sec:discussion}

Observing the 3D magnetic fields of molecular clouds and their substructures is essential for understanding their formation mechanism and the role magnetic fields play in star formation. Observations of \blos\ and \bperp\ are necessary but insufficient for determining the 3D fields. While \blos\ observing techniques provide both the strength and direction of this component, \bperp\ observing techniques provide only the orientation and strength of this component, but not its direction. 
Knowing the strengths and complete directions of \blos\ and \bperp\   enables us to infer the ordered, line-of-sight-averaged 3D field vectors. However, due to systematic biases between the techniques for determining field strengths, additional observations, such as observing the magnetic field inclination angles are required. 
The \blos\ strength and direction, $\gamma$, and \bperp\ orientation (without its direction) do not fully infer the 3D fields, as they can lead to two different vectors depicted in Figure~\ref{fig:inclination}.  Other techniques such as the use of GMF models~\citep{Tahanietal2022O, Tahanietal2022P} can help resolve this issue. 

The studies of \blos , \bperp , $\gamma$, and GMF could enable us to infer the 3D ordered magnetic fields of molecular clouds with improved precision. Upcoming observations will 1) enhance the precision and accuracy of the inferred 3D magnetic field of each cloud,  2) result in 3D magnetic field maps of more regions, \ch{and 3) produce more accurate GMF models, thereby enhancing the technique's underlying assumptions.} 

The forthcoming Zeeman measurements~\citep[][for the most accurate determination of field strengths]{Robishawetal2015} and Faraday rotation measure catalogs by the Square Kilometer Array (SKA) project~\citep{Healdetal2020} or the Australian Square Kilometer Array Pathfinder (ASKAP), such as the Polarisation Sky Survey of the Universe's Magnetism (POSSUM) rotation measure catalog~\citep[]{Gaensler2010}, will provide the \blos\ of numerous molecular clouds with lower uncertainties and greater source density than previous catalogs~\citep[e.g.,][]{Tayloretal2009}. These observations will increase the number of \blos\ detections per molecular cloud by a factor of $\sim 10$. These \blos\ maps and future \bperp\ observations, such as those by the Fred Young Sub-millimeter Telescope~\citep[FYST;][]{CCATPrime2021}, 
will enable 3D magnetic field maps of many molecular clouds.

Finally, starlight polarization observations~\citep[e.g.,][]{PereyraMagalhaes2007} combined with Gaia-observed parallax distances allow us to  differentiate between, and separate, various cloud components along the line of sight~\citep[e.g.,][]{Doietal2021}. This is made possible by existing and upcoming starlight polarization observations, including the Galactic Plane Infrared Polarization Survey~\citep[GPIPS;][]{Clemensetal2020} and the upcoming optical polarimetry survey with the Polar-Areas Stellar Imaging Polarization High Accuracy Experiment~\citep[PASIPHAE;][]{Tassisetal2018}.

\section*{Acknowledgments}
We appreciate the referees' insightful, thorough, and diligent comments. We thank Huirong Yan for helpful conversation about atomic alignment.  Figure~\ref{fig:Blos3D}
employs a function written by Susan Clark and later modified by Jennifer Glover~\citep{Tahanietal2022O} to perform line integration convolution. Quillbot was utilized for editing purposes. MT was hired by the National Research Council Canada. MT is supported by the Banting Fellowship (Natural Sciences and Engineering Research Council Canada) hosted at Stanford University and the Kavli Institute for Particle Astrophysics and Cosmology (KIPAC) Fellowship.

\bibliography{AllBiblio}{}

\begin{thebibliography}{145}
\providecommand{\natexlab}[1]{#1}
\expandafter\ifx\csname urlstyle\endcsname\relax
  \providecommand{\doi}[1]{doi:\discretionary{}{}{}#1}\else
  \providecommand{\doi}{doi:\discretionary{}{}{}\begingroup
  \urlstyle{rm}\Url}\fi
\providecommand{\selectlanguage}[1]{\relax}
\providecommand{\bibAnnoteFile}[1]{%
  \IfFileExists{#1}{\begin{quotation}\noindent\textsc{Key:} #1\\
  \textsc{Annotation:}\ \input{#1}\end{quotation}}{}}
\providecommand{\bibAnnote}[2]{%
  \begin{quotation}\noindent\textsc{Key:} #1\\
  \textsc{Annotation:}\ #2\end{quotation}}

\bibitem[{{Abe} et~al.(2021){Abe}, {Inoue}, {Inutsuka}, and
  {Matsumoto}}]{Abeetal2021}
{Abe}, D., {Inoue}, T., {Inutsuka}, S.-i., and {Matsumoto}, T. (2021).
\newblock {Classification of Filament Formation Mechanisms in Magnetized
  Molecular Clouds}.
\newblock \emph{\apj} 916, 83.
\newblock \doi{10.3847/1538-4357/ac07a1}
\bibAnnoteFile{Abeetal2021}

\bibitem[{{Alina} et~al.(2019){Alina}, {Ristorcelli}, {Montier}, {Abdikamalov},
  {Juvela}, {Ferri{\`e}re} et~al.}]{Alinaetal2019}
{Alina}, D., {Ristorcelli}, I., {Montier}, L., {Abdikamalov}, E., {Juvela}, M.,
  {Ferri{\`e}re}, K., et~al. (2019).
\newblock {Statistical analysis of the interplay between interstellar magnetic
  fields and filaments hosting Planck Galactic cold clumps}.
\newblock \emph{\mnras} 485, 2825--2843.
\newblock \doi{10.1093/mnras/stz508}
\bibAnnoteFile{Alinaetal2019}

\bibitem[{{Alves} et~al.(2018){Alves}, {Boulanger}, {Ferri{\`e}re}, and
  {Montier}}]{Alvesetal2018}
{Alves}, M.~I.~R., {Boulanger}, F., {Ferri{\`e}re}, K., and {Montier}, L.
  (2018).
\newblock {The Local Bubble: a magnetic veil to our Galaxy}.
\newblock \emph{\aap} 611, L5.
\newblock \doi{10.1051/0004-6361/201832637}
\bibAnnoteFile{Alvesetal2018}

\bibitem[{{Andersson} et~al.(2015){Andersson}, {Lazarian}, and
  {Vaillancourt}}]{Anderssonetal2015}
{Andersson}, B.-G., {Lazarian}, A., and {Vaillancourt}, J.~E. (2015).
\newblock {Interstellar Dust Grain Alignment}.
\newblock \emph{\araa} 53, 501--539.
\newblock \doi{10.1146/annurev-astro-082214-122414}
\bibAnnoteFile{Anderssonetal2015}

\bibitem[{{Arzoumanian} et~al.(2021){Arzoumanian}, {Furuya}, {Hasegawa},
  {Tahani}, {Sadavoy}, {Hull} et~al.}]{Arzoumanianetal2021}
{Arzoumanian}, D., {Furuya}, R.~S., {Hasegawa}, T., {Tahani}, M., {Sadavoy},
  S., {Hull}, C.~L.~H., et~al. (2021).
\newblock {Dust polarized emission observations of NGC 6334. BISTRO reveals the
  details of the complex but organized magnetic field structure of the
  high-mass star-forming hub-filament network}.
\newblock \emph{\aap} 647, A78.
\newblock \doi{10.1051/0004-6361/202038624}
\bibAnnoteFile{Arzoumanianetal2021}

\bibitem[{{Arzoumanian} et~al.(2018){Arzoumanian}, {Shimajiri}, {Inutsuka},
  {Inoue}, and {Tachihara}}]{Arzoumanianetal2018}
{Arzoumanian}, D., {Shimajiri}, Y., {Inutsuka}, S.-i., {Inoue}, T., and
  {Tachihara}, K. (2018).
\newblock {Molecular filament formation and filament-cloud interaction: Hints
  from Nobeyama 45 m telescope observations}.
\newblock \emph{\pasj} 70, 96.
\newblock \doi{10.1093/pasj/psy095}
\bibAnnoteFile{Arzoumanianetal2018}

\bibitem[{{Beck}(2015)}]{Beck2015}
{Beck}, R. (2015).
\newblock {Magnetic fields in spiral galaxies}.
\newblock \emph{\aapr} 24, 4.
\newblock \doi{10.1007/s00159-015-0084-4}
\bibAnnoteFile{Beck2015}

\bibitem[{{Bialy} et~al.(2021){Bialy}, {Zucker}, {Goodman}, {Foley}, {Alves},
  {Semenov} et~al.}]{Bialy2021}
{Bialy}, S., {Zucker}, C., {Goodman}, A., {Foley}, M.~M., {Alves}, J.,
  {Semenov}, V.~A., et~al. (2021).
\newblock {The Per-Tau Shell: A Giant Star-forming Spherical Shell Revealed by
  3D Dust Observations}.
\newblock \emph{\apjl} 919, L5.
\newblock \doi{10.3847/2041-8213/ac1f95}
\bibAnnoteFile{Bialy2021}

\bibitem[{{Bonne} et~al.(2020){Bonne}, {Bontemps}, {Schneider}, {Clarke},
  {Arzoumanian}, {Fukui} et~al.}]{Bonneetal2020}
{Bonne}, L., {Bontemps}, S., {Schneider}, N., {Clarke}, S.~D., {Arzoumanian},
  D., {Fukui}, Y., et~al. (2020).
\newblock {Formation of the Musca filament: evidence for asymmetries in the
  accretion flow due to a cloud-cloud collision}.
\newblock \emph{\aap} 644, A27.
\newblock \doi{10.1051/0004-6361/202038281}
\bibAnnoteFile{Bonneetal2020}

\bibitem[{{Bracco} et~al.(2020){Bracco}, {Bresnahan}, {Palmeirim},
  {Arzoumanian}, {Andr{\'e}}, {Ward-Thompson} et~al.}]{Braccoetal2020}
{Bracco}, A., {Bresnahan}, D., {Palmeirim}, P., {Arzoumanian}, D., {Andr{\'e}},
  P., {Ward-Thompson}, D., et~al. (2020).
\newblock {Compressed magnetized shells of atomic gas and the formation of the
  Corona Australis molecular cloud}.
\newblock \emph{\aap} 644, A5.
\newblock \doi{10.1051/0004-6361/202039282}
\bibAnnoteFile{Braccoetal2020}

\bibitem[{{Brown} et~al.(2008){Brown}, {Stil}, and {Landecker}}]{Brown2008}
{Brown}, J.~C., {Stil}, J.~M., and {Landecker}, T.~L. (2008).
\newblock {Visualizing the Invisible using Polarisation Observations}.
\newblock \emph{Physics in Canada} 64
\bibAnnoteFile{Brown2008}

\bibitem[{{CCAT-Prime collaboration} et~al.(2021){CCAT-Prime collaboration},
  {Aravena}, {Austermann}, {Basu}, {Battaglia}, {Beringue}
  et~al.}]{CCATPrime2021}
{CCAT-Prime collaboration}, {Aravena}, M., {Austermann}, J.~E., {Basu}, K.,
  {Battaglia}, N., {Beringue}, B., et~al. (2021).
\newblock {CCAT-prime Collaboration: Science Goals and Forecasts with Prime-Cam
  on the Fred Young Submillimeter Telescope}.
\newblock \emph{arXiv e-prints} , arXiv:2107.10364
\bibAnnoteFile{CCATPrime2021}

\bibitem[{{Chandrasekhar} and {Fermi}(1953)}]{ChandrasekharFermi1953}
{Chandrasekhar}, S. and {Fermi}, E. (1953).
\newblock {Magnetic Fields in Spiral Arms.}
\newblock \emph{\apj} 118, 113.
\newblock \doi{10.1086/145731}
\bibAnnoteFile{ChandrasekharFermi1953}

\bibitem[{{Chen} et~al.(2016){Chen}, {King}, and {Li}}]{Chenetal2016}
{Chen}, C.-Y., {King}, P.~K., and {Li}, Z.-Y. (2016).
\newblock {Change of Magnetic Field-gas Alignment at the Gravity-driven
  Alfv{\'e}nic Transition in Molecular Clouds: Implications for Dust
  Polarization Observations}.
\newblock \emph{\apj} 829, 84.
\newblock \doi{10.3847/0004-637X/829/2/84}
\bibAnnoteFile{Chenetal2016}

\bibitem[{{Chen} et~al.(2019){Chen}, {King}, {Li}, {Fissel}, and
  {Mazzei}}]{Chenetal2019}
{Chen}, C.-Y., {King}, P.~K., {Li}, Z.-Y., {Fissel}, L.~M., and {Mazzei}, R.~R.
  (2019).
\newblock {A new method to trace three-dimensional magnetic field structure
  within molecular clouds using dust polarization}.
\newblock \emph{\mnras} 485, 3499--3513.
\newblock \doi{10.1093/mnras/stz618}
\bibAnnoteFile{Chenetal2019}

\bibitem[{{Chuss} et~al.(2019){Chuss}, {Andersson}, {Bally}, {Dotson},
  {Dowell}, {Guerra} et~al.}]{Chussetal2019}
{Chuss}, D.~T., {Andersson}, B.~G., {Bally}, J., {Dotson}, J.~L., {Dowell},
  C.~D., {Guerra}, J.~A., et~al. (2019).
\newblock {HAWC+/SOFIA Multiwavelength Polarimetric Observations of OMC-1}.
\newblock \emph{\apj} 872, 187.
\newblock \doi{10.3847/1538-4357/aafd37}
\bibAnnoteFile{Chussetal2019}

\bibitem[{{Clark}(2018)}]{Clark2018}
{Clark}, S.~E. (2018).
\newblock {A New Probe of Line-of-sight Magnetic Field Tangling}.
\newblock \emph{\apjl} 857, L10.
\newblock \doi{10.3847/2041-8213/aabb54}
\bibAnnoteFile{Clark2018}

\bibitem[{{Clark} and {Hensley}(2019)}]{ClarkHensley2019}
{Clark}, S.~E. and {Hensley}, B.~S. (2019).
\newblock {Mapping the Magnetic Interstellar Medium in Three Dimensions over
  the Full Sky with Neutral Hydrogen}.
\newblock \emph{\apj} 887, 136.
\newblock \doi{10.3847/1538-4357/ab5803}
\bibAnnoteFile{ClarkHensley2019}

\bibitem[{{Clark} et~al.(2015){Clark}, {Hill}, {Peek}, {Putman}, and
  {Babler}}]{Clarketal2015}
{Clark}, S.~E., {Hill}, J.~C., {Peek}, J.~E.~G., {Putman}, M.~E., and {Babler},
  B.~L. (2015).
\newblock {Neutral Hydrogen Structures Trace Dust Polarization Angle:
  Implications for Cosmic Microwave Background Foregrounds}.
\newblock \emph{\prl} 115, 241302.
\newblock \doi{10.1103/PhysRevLett.115.241302}
\bibAnnoteFile{Clarketal2015}

\bibitem[{{Clark} et~al.(2014){Clark}, {Peek}, and {Putman}}]{Clarketal2014}
{Clark}, S.~E., {Peek}, J.~E.~G., and {Putman}, M.~E. (2014).
\newblock {Magnetically Aligned H I Fibers and the Rolling Hough Transform}.
\newblock \emph{\apj} 789, 82.
\newblock \doi{10.1088/0004-637X/789/1/82}
\bibAnnoteFile{Clarketal2014}

\bibitem[{{Clemens} et~al.(2020){Clemens}, {Cashman}, {Cerny}, {El-Batal},
  {Jameson}, {Marchwinski} et~al.}]{Clemensetal2020}
{Clemens}, D.~P., {Cashman}, L.~R., {Cerny}, C., {El-Batal}, A.~M., {Jameson},
  K.~E., {Marchwinski}, R., et~al. (2020).
\newblock {The Galactic Plane Infrared Polarization Survey (GPIPS): Data
  Release 4}.
\newblock \emph{\apjs} 249, 23.
\newblock \doi{10.3847/1538-4365/ab9f30}
\bibAnnoteFile{Clemensetal2020}

\bibitem[{{Clemens} et~al.(2016){Clemens}, {Tassis}, and
  {Goldsmith}}]{Clemensetal2016}
{Clemens}, D.~P., {Tassis}, K., and {Goldsmith}, P.~F. (2016).
\newblock {The Magnetic Field of L1544. I. Near-infrared Polarimetry and the
  Non-uniform Envelope}.
\newblock \emph{\apj} 833, 176.
\newblock \doi{10.3847/1538-4357/833/2/176}
\bibAnnoteFile{Clemensetal2016}

\bibitem[{{Cort{\'e}s} et~al.(2021){Cort{\'e}s}, {Sanhueza}, {Houde},
  {Mart{\'\i}n}, {Hull}, {Girart} et~al.}]{Cortesetal2021}
{Cort{\'e}s}, P.~C., {Sanhueza}, P., {Houde}, M., {Mart{\'\i}n}, S., {Hull}, C.
  L.~H., {Girart}, J.~M., et~al. (2021).
\newblock {Magnetic Fields in Massive Star-forming Regions (MagMaR). II.
  Tomography through Dust and Molecular Line Polarization in NGC 6334I(N)}.
\newblock \emph{\apj} 923, 204.
\newblock \doi{10.3847/1538-4357/ac28a1}
\bibAnnoteFile{Cortesetal2021}

\bibitem[{{Crutcher} and {Kemball}(2019)}]{CrutcherKemball2019}
{Crutcher}, R.~M. and {Kemball}, A.~J. (2019).
\newblock {Review of Zeeman Effect Observations of Regions of Star Formation K
  Zeeman Effect, Magnetic Fields, Star formation, Masers, Molecular clouds}.
\newblock \emph{Frontiers in Astronomy and Space Sciences} 6, 66.
\newblock \doi{10.3389/fspas.2019.00066}
\bibAnnoteFile{CrutcherKemball2019}

\bibitem[{{Crutcher} et~al.(2010){Crutcher}, {Wandelt}, {Heiles}, {Falgarone},
  and {Troland}}]{Crutcheretal2010}
{Crutcher}, R.~M., {Wandelt}, B., {Heiles}, C., {Falgarone}, E., and {Troland},
  T.~H. (2010).
\newblock {Magnetic Fields in Interstellar Clouds from Zeeman Observations:
  Inference of Total Field Strengths by Bayesian Analysis}.
\newblock \emph{\apj} 725, 466--479.
\newblock \doi{10.1088/0004-637X/725/1/466}
\bibAnnoteFile{Crutcheretal2010}

\bibitem[{{Davis} and {Greenstein}(1951)}]{DavisGreenstein1951}
{Davis}, L.~J. and {Greenstein}, J.~L. (1951).
\newblock {The Polarization of Starlight by Aligned Dust Grains.}
\newblock \emph{\apj} 114, 206.
\newblock \doi{10.1086/145464}
\bibAnnoteFile{DavisGreenstein1951}

\bibitem[{{Doi} et~al.(2021){Doi}, {Hasegawa}, {Bastien}, {Tahani},
  {Arzoumanian}, {Coud{\'e}} et~al.}]{Doietal2021}
{Doi}, Y., {Hasegawa}, T., {Bastien}, P., {Tahani}, M., {Arzoumanian}, D.,
  {Coud{\'e}}, S., et~al. (2021).
\newblock {Two-component Magnetic Field along the Line of Sight to the Perseus
  Molecular Cloud: Contribution of the Foreground Taurus Molecular Cloud}.
\newblock \emph{\apj} 914, 122.
\newblock \doi{10.3847/1538-4357/abfcc5}
\bibAnnoteFile{Doietal2021}

\bibitem[{{Draine}(2003)}]{Draine2003}
{Draine}, B.~T. (2003).
\newblock {Interstellar Dust Grains}.
\newblock \emph{\araa} 41, 241--289.
\newblock \doi{10.1146/annurev.astro.41.011802.094840}
\bibAnnoteFile{Draine2003}

\bibitem[{{Draine}(2009)}]{Draine2009}
{Draine}, B.~T. (2009).
\newblock {Interstellar Dust Models and Evolutionary Implications}.
\newblock In \emph{Cosmic Dust - Near and Far}, eds. T.~{Henning},
  E.~{Gr{\"u}n}, and J.~{Steinacker}. vol. 414 of \emph{Astronomical Society of
  the Pacific Conference Series}, 453
\bibAnnoteFile{Draine2009}

\bibitem[{{Draine} and {Weingartner}(1997)}]{Draineetal1997}
{Draine}, B.~T. and {Weingartner}, J.~C. (1997).
\newblock {Radiative Torques on Interstellar Grains. II. Grain Alignment}.
\newblock \emph{\apj} 480, 633--646.
\newblock \doi{10.1086/304008}
\bibAnnoteFile{Draineetal1997}

\bibitem[{{Eswaraiah} et~al.(2021){Eswaraiah}, {Li}, {Furuya}, {Hasegawa},
  {Ward-Thompson}, {Qiu} et~al.}]{Eswaraiahetal2021}
{Eswaraiah}, C., {Li}, D., {Furuya}, R.~S., {Hasegawa}, T., {Ward-Thompson},
  D., {Qiu}, K., et~al. (2021).
\newblock {The JCMT BISTRO Survey: Revealing the Diverse Magnetic Field
  Morphologies in Taurus Dense Cores with Sensitive Submillimeter Polarimetry}.
\newblock \emph{\apjl} 912, L27.
\newblock \doi{10.3847/2041-8213/abeb1c}
\bibAnnoteFile{Eswaraiahetal2021}

\bibitem[{{Falceta-Gon{\c{c}}alves} et~al.(2008){Falceta-Gon{\c{c}}alves},
  {Lazarian}, and {Kowal}}]{FalcetaGoncalvesetal2008}
{Falceta-Gon{\c{c}}alves}, D., {Lazarian}, A., and {Kowal}, G. (2008).
\newblock {Studies of Regular and Random Magnetic Fields in the ISM: Statistics
  of Polarization Vectors and the Chandrasekhar-Fermi Technique}.
\newblock \emph{\apj} 679, 537--551.
\newblock \doi{10.1086/587479}
\bibAnnoteFile{FalcetaGoncalvesetal2008}

\bibitem[{{Fauvet} et~al.(2011){Fauvet}, {Mac{\'\i}as-P{\'e}rez}, {Aumont},
  {D{\'e}sert}, {Jaffe}, {Banday} et~al.}]{Fauvetetal2011}
{Fauvet}, L., {Mac{\'\i}as-P{\'e}rez}, J.~F., {Aumont}, J., {D{\'e}sert},
  F.~X., {Jaffe}, T.~R., {Banday}, A.~J., et~al. (2011).
\newblock {Joint 3D modelling of the polarized Galactic synchrotron and thermal
  dust foreground diffuse emission}.
\newblock \emph{\aap} 526, A145.
\newblock \doi{10.1051/0004-6361/201014492}
\bibAnnoteFile{Fauvetetal2011}

\bibitem[{{Ferri{\`e}re}(2016)}]{Ferriere2016}
{Ferri{\`e}re}, K. (2016).
\newblock {Faraday tomography: a new, three-dimensional probe of the
  interstellar magnetic field}.
\newblock In \emph{Journal of Physics Conference Series}. vol. 767 of
  \emph{Journal of Physics Conference Series}, 012006.
\newblock \doi{10.1088/1742-6596/767/1/012006}
\bibAnnoteFile{Ferriere2016}

\bibitem[{{Fiege} and {Pudritz}(2000{\natexlab{a}})}]{FiegePudritzI2000}
{Fiege}, J.~D. and {Pudritz}, R.~E. (2000{\natexlab{a}}).
\newblock {Helical fields and filamentary molecular clouds - I}.
\newblock \emph{\mnras} 311, 85--104.
\newblock \doi{10.1046/j.1365-8711.2000.03066.x}
\bibAnnoteFile{FiegePudritzI2000}

\bibitem[{{Fiege} and {Pudritz}(2000{\natexlab{b}})}]{FiegePudritzII2000}
{Fiege}, J.~D. and {Pudritz}, R.~E. (2000{\natexlab{b}}).
\newblock {Helical fields and filamentary molecular clouds - II. Axisymmetric
  stability and fragmentation}.
\newblock \emph{\mnras} 311, 105--119.
\newblock \doi{10.1046/j.1365-8711.2000.03067.x}
\bibAnnoteFile{FiegePudritzII2000}

\bibitem[{{Fissel} et~al.(2019){Fissel}, {Ade}, {Angil{\`e}}, {Ashton},
  {Benton}, {Chen} et~al.}]{Fisseletal2019}
{Fissel}, L.~M., {Ade}, P. A.~R., {Angil{\`e}}, F.~E., {Ashton}, P., {Benton},
  S.~J., {Chen}, C.-Y., et~al. (2019).
\newblock {Relative Alignment between the Magnetic Field and Molecular Gas
  Structure in the Vela C Giant Molecular Cloud Using Low- and High-density
  Tracers}.
\newblock \emph{\apj} 878, 110.
\newblock \doi{10.3847/1538-4357/ab1eb0}
\bibAnnoteFile{Fisseletal2019}

\bibitem[{{Fissel} et~al.(2016){Fissel}, {Ade}, {Angil{\`e}}, {Ashton},
  {Benton}, {Devlin} et~al.}]{Fisseletal2016}
{Fissel}, L.~M., {Ade}, P. A.~R., {Angil{\`e}}, F.~E., {Ashton}, P., {Benton},
  S.~J., {Devlin}, M.~J., et~al. (2016).
\newblock {Balloon-Borne Submillimeter Polarimetry of the Vela C Molecular
  Cloud: Systematic Dependence of Polarization Fraction on Column Density and
  Local Polarization-Angle Dispersion}.
\newblock \emph{\apj} 824, 134.
\newblock \doi{10.3847/0004-637X/824/2/134}
\bibAnnoteFile{Fisseletal2016}

\bibitem[{{Gaensler} et~al.(2010){Gaensler}, {Landecker}, {Taylor}, and {POSSUM
  Collaboration}}]{Gaensler2010}
{Gaensler}, B.~M., {Landecker}, T.~L., {Taylor}, A.~R., and {POSSUM
  Collaboration} (2010).
\newblock {Survey Science with ASKAP: Polarization Sky Survey of the Universe's
  Magnetism (POSSUM)}.
\newblock In \emph{American Astronomical Society Meeting Abstracts \#215}. vol.
  215 of \emph{American Astronomical Society Meeting Abstracts}, 470.13
\bibAnnoteFile{Gaensler2010}

\bibitem[{{Gaia Collaboration} et~al.(2016){Gaia Collaboration}, {Prusti}, {de
  Bruijne}, {Brown}, {Vallenari}, {Babusiaux} et~al.}]{GaiaMission2016}
{Gaia Collaboration}, {Prusti}, T., {de Bruijne}, J.~H.~J., {Brown}, A.~G.~A.,
  {Vallenari}, A., {Babusiaux}, C., et~al. (2016).
\newblock {The Gaia mission}.
\newblock \emph{\aap} 595, A1.
\newblock \doi{10.1051/0004-6361/201629272}
\bibAnnoteFile{GaiaMission2016}

\bibitem[{{Gibson} et~al.(2009){Gibson}, {Plume}, {Bergin}, {Ragan}, and
  {Evans}}]{Gibsonetal2009}
{Gibson}, D., {Plume}, R., {Bergin}, E., {Ragan}, S., and {Evans}, N. (2009).
\newblock {Molecular Line Observations of Infrared Dark Clouds. II. Physical
  Conditions}.
\newblock \emph{\apj} 705, 123--134.
\newblock \doi{10.1088/0004-637X/705/1/123}
\bibAnnoteFile{Gibsonetal2009}

\bibitem[{{Girichidis}(2021)}]{Girichidis2021}
{Girichidis}, P. (2021).
\newblock {Alignment of the magnetic field in star-forming regions and why it
  might be difficult to observe}.
\newblock \emph{\mnras} 507, 5641--5657.
\newblock \doi{10.1093/mnras/stab2157}
\bibAnnoteFile{Girichidis2021}

\bibitem[{{G{\'o}mez} et~al.(2018){G{\'o}mez}, {V{\'a}zquez-Semadeni}, and
  {Zamora-Avil{\'e}s}}]{Gomezetal2018}
{G{\'o}mez}, G.~C., {V{\'a}zquez-Semadeni}, E., and {Zamora-Avil{\'e}s}, M.
  (2018).
\newblock {The magnetic field structure in molecular cloud filaments}.
\newblock \emph{\mnras} 480, 2939--2944.
\newblock \doi{10.1093/mnras/sty2018}
\bibAnnoteFile{Gomezetal2018}

\bibitem[{{Gonz{\'a}lez-Casanova} and
  {Lazarian}(2017)}]{GonzalezCasanovaLazarian2017}
{Gonz{\'a}lez-Casanova}, D.~F. and {Lazarian}, A. (2017).
\newblock {Velocity Gradients as a Tracer for Magnetic Fields}.
\newblock \emph{\apj} 835, 41.
\newblock \doi{10.3847/1538-4357/835/1/41}
\bibAnnoteFile{GonzalezCasanovaLazarian2017}

\bibitem[{{Gro{\ss}schedl} et~al.(2018){Gro{\ss}schedl}, {Alves}, {Meingast},
  {Ackerl}, {Ascenso}, {Bouy} et~al.}]{Grossschedletal2018}
{Gro{\ss}schedl}, J.~E., {Alves}, J., {Meingast}, S., {Ackerl}, C., {Ascenso},
  J., {Bouy}, H., et~al. (2018).
\newblock {3D shape of Orion A from Gaia DR2}.
\newblock \emph{\aap} 619, A106.
\newblock \doi{10.1051/0004-6361/201833901}
\bibAnnoteFile{Grossschedletal2018}

\bibitem[{{Hacar} et~al.(2022){Hacar}, {Clark}, {Heitsch}, {Kainulainen},
  {Panopoulou}, {Seifried} et~al.}]{Hacaretal2022}
{Hacar}, A., {Clark}, S., {Heitsch}, F., {Kainulainen}, J., {Panopoulou}, G.,
  {Seifried}, D., et~al. (2022).
\newblock {Initial Conditions for Star Formation: A Physical Description of the
  Filamentary ISM}.
\newblock \emph{arXiv e-prints} , arXiv:2203.09562
\bibAnnoteFile{Hacaretal2022}

\bibitem[{{Han}(2017)}]{Han2017}
{Han}, J.~L. (2017).
\newblock {Observing Interstellar and Intergalactic Magnetic Fields}.
\newblock \emph{\araa} 55, 111--157.
\newblock \doi{10.1146/annurev-astro-091916-055221}
\bibAnnoteFile{Han2017}

\bibitem[{{Han} and {Zhang}(2007)}]{HanZhang2007}
{Han}, J.~L. and {Zhang}, J.~S. (2007).
\newblock {The Galactic distribution of magnetic fields in molecular clouds and
  HII regions}.
\newblock \emph{\aap} 464, 609--614.
\newblock \doi{10.1051/0004-6361:20065801}
\bibAnnoteFile{HanZhang2007}

\bibitem[{{Haverkorn}(2015)}]{Haverkorn2015}
{Haverkorn}, M. (2015).
\newblock {Magnetic Fields in the Milky Way}.
\newblock In \emph{Magnetic Fields in Diffuse Media}, eds. A.~{Lazarian}, E.~M.
  {de Gouveia Dal Pino}, and C.~{Melioli}. vol. 407 of \emph{Astrophysics and
  Space Science Library}, 483.
\newblock \doi{10.1007/978-3-662-44625-6\_17}
\bibAnnoteFile{Haverkorn2015}

\bibitem[{{Heald} et~al.(2020){Heald}, {Mao}, {Vacca}, {Akahori},
  {Damas-Segovia}, {Gaensler} et~al.}]{Healdetal2020}
{Heald}, G., {Mao}, S., {Vacca}, V., {Akahori}, T., {Damas-Segovia}, A.,
  {Gaensler}, B., et~al. (2020).
\newblock {Magnetism Science with the Square Kilometre Array}.
\newblock \emph{Galaxies} 8, 53.
\newblock \doi{10.3390/galaxies8030053}
\bibAnnoteFile{Healdetal2020}

\bibitem[{{Heiles}(1997)}]{Heiles1997}
{Heiles}, C. (1997).
\newblock {A Holistic View of the Magnetic Field in the Eridanus/Orion Region}.
\newblock \emph{\apjs} 111, 245--288.
\newblock \doi{10.1086/313010}
\bibAnnoteFile{Heiles1997}

\bibitem[{{Hennebelle} and {Inutsuka}(2019)}]{HennebelleInutsuka2019}
{Hennebelle}, P. and {Inutsuka}, S.-i. (2019).
\newblock {The role of magnetic field in molecular cloud formation and
  evolution}.
\newblock \emph{arXiv e-prints} , arXiv:1902.00798
\bibAnnoteFile{HennebelleInutsuka2019}

\bibitem[{{Hensley} et~al.(2019){Hensley}, {Zhang}, and
  {Bock}}]{Hensleyetal2019}
{Hensley}, B.~S., {Zhang}, C., and {Bock}, J.~J. (2019).
\newblock {An Imprint of the Galactic Magnetic Field in the Diffuse Unpolarized
  Dust Emission}.
\newblock \emph{\apj} 887, 159.
\newblock \doi{10.3847/1538-4357/ab5183}
\bibAnnoteFile{Hensleyetal2019}

\bibitem[{{Hezareh} et~al.(2010){Hezareh}, {Houde}, {McCoey}, and
  {Li}}]{Hezarehetal2010}
{Hezareh}, T., {Houde}, M., {McCoey}, C., and {Li}, H.-b. (2010).
\newblock {Observational Determination of the Turbulent Ambipolar Diffusion
  Scale and Magnetic Field Strength in Molecular Clouds}.
\newblock \emph{\apj} 720, 603--607.
\newblock \doi{10.1088/0004-637X/720/1/603}
\bibAnnoteFile{Hezarehetal2010}

\bibitem[{{Hoang} and {Lazarian}(2016)}]{HoangLazarian2016}
{Hoang}, T. and {Lazarian}, A. (2016).
\newblock {A Unified Model of Grain Alignment: Radiative Alignment of
  Interstellar Grains with Magnetic Inclusions}.
\newblock \emph{\apj} 831, 159.
\newblock \doi{10.3847/0004-637X/831/2/159}
\bibAnnoteFile{HoangLazarian2016}

\bibitem[{{Houde}(2011)}]{Houde2011}
{Houde}, M. (2011).
\newblock {Magnetic Fields in Three Dimensions}.
\newblock In \emph{Astronomical Polarimetry 2008: Science from Small to Large
  Telescopes}, eds. P.~{Bastien}, N.~{Manset}, D.~P. {Clemens}, and
  N.~{St-Louis}. vol. 449 of \emph{Astronomical Society of the Pacific
  Conference Series}, 213
\bibAnnoteFile{Houde2011}

\bibitem[{{Houde} et~al.(2002){Houde}, {Bastien}, {Dotson}, {Dowell},
  {Hildebrand}, {Peng} et~al.}]{Houdeetal2002}
{Houde}, M., {Bastien}, P., {Dotson}, J.~L., {Dowell}, C.~D., {Hildebrand},
  R.~H., {Peng}, R., et~al. (2002).
\newblock {On the Measurement of the Magnitude and Orientation of the Magnetic
  Field in Molecular Clouds}.
\newblock \emph{\apj} 569, 803--814.
\newblock \doi{10.1086/339356}
\bibAnnoteFile{Houdeetal2002}

\bibitem[{{Houde} et~al.(2000{\natexlab{a}}){Houde}, {Bastien}, {Peng},
  {Phillips}, and {Yoshida}}]{Houdeetal2000a}
{Houde}, M., {Bastien}, P., {Peng}, R., {Phillips}, T.~G., and {Yoshida}, H.
  (2000{\natexlab{a}}).
\newblock {Probing the Magnetic Field with Molecular Ion Spectra}.
\newblock \emph{\apj} 536, 857--864.
\newblock \doi{10.1086/308980}
\bibAnnoteFile{Houdeetal2000a}

\bibitem[{{Houde} et~al.(2000{\natexlab{b}}){Houde}, {Peng}, {Phillips},
  {Bastien}, and {Yoshida}}]{Houdeetal2000b}
{Houde}, M., {Peng}, R., {Phillips}, T.~G., {Bastien}, P., and {Yoshida}, H.
  (2000{\natexlab{b}}).
\newblock {Probing the Magnetic Field with Molecular Ion Spectra. II.}
\newblock \emph{\apj} 537, 245--254.
\newblock \doi{10.1086/309035}
\bibAnnoteFile{Houdeetal2000b}

\bibitem[{{Houde} et~al.(2004){Houde}, {Peng}, {Yoshida}, {Hildebrand},
  {Phillips}, {Dowell} et~al.}]{Houdeetal2004}
{Houde}, M., {Peng}, R., {Yoshida}, H., {Hildebrand}, R.~H., {Phillips}, T.~G.,
  {Dowell}, C.~D., et~al. (2004).
\newblock {The Measurement of the Orientation of the Magnetic Field in
  Molecular Clouds}.
\newblock \emph{\apss} 292, 127--134.
\newblock \doi{10.1023/B:ASTR.0000045008.39439.5b}
\bibAnnoteFile{Houdeetal2004}

\bibitem[{{Houde} et~al.(2009){Houde}, {Vaillancourt}, {Hildebrand},
  {Chitsazzadeh}, and {Kirby}}]{Houdeetal2009}
{Houde}, M., {Vaillancourt}, J.~E., {Hildebrand}, R.~H., {Chitsazzadeh}, S.,
  and {Kirby}, L. (2009).
\newblock {Dispersion of Magnetic Fields in Molecular Clouds. II.}
\newblock \emph{\apj} 706, 1504--1516.
\newblock \doi{10.1088/0004-637X/706/2/1504}
\bibAnnoteFile{Houdeetal2009}

\bibitem[{{Hu} and {Lazarian}(2022)}]{HuLazarian2022}
{Hu}, Y. and {Lazarian}, A. (2022).
\newblock {Probing Three-Dimensional Magnetic Fields: I -- Polarized Dust
  Emission}.
\newblock \emph{arXiv e-prints} , arXiv:2203.09745
\bibAnnoteFile{HuLazarian2022}

\bibitem[{{Hu} et~al.(2022){Hu}, {Lazarian}, and {Wang}}]{HuLazarianWang2022}
{Hu}, Y., {Lazarian}, A., and {Wang}, Q.~D. (2022).
\newblock {Decomposing Magnetic Fields in Three Dimensions over the Central
  Molecular Zone}.
\newblock \emph{\mnras} \doi{10.1093/mnras/stac1060}
\bibAnnoteFile{HuLazarianWang2022}

\bibitem[{{Hu} et~al.(2021{\natexlab{a}}){Hu}, {Lazarian}, and
  {Xu}}]{Huetal2021PPV}
{Hu}, Y., {Lazarian}, A., and {Xu}, S. (2021{\natexlab{a}}).
\newblock {Anisotropic Turbulence in Position-Position-Velocity Space: Probing
  Three-dimensional Magnetic Fields}.
\newblock \emph{\apj} 915, 67.
\newblock \doi{10.3847/1538-4357/ac00ab}
\bibAnnoteFile{Huetal2021PPV}

\bibitem[{{Hu} et~al.(2020){Hu}, {Lazarian}, and {Yuen}}]{Huetal2020}
{Hu}, Y., {Lazarian}, A., and {Yuen}, K.~H. (2020).
\newblock {Velocity Gradient in the Presence of Self-gravity: Identifying
  Gravity-induced Inflow and Determining Collapsing Stage}.
\newblock \emph{\apj} 897, 123.
\newblock \doi{10.3847/1538-4357/ab9948}
\bibAnnoteFile{Huetal2020}

\bibitem[{{Hu} et~al.(2021{\natexlab{b}}){Hu}, {Xu}, and
  {Lazarian}}]{Huetal2021}
{Hu}, Y., {Xu}, S., and {Lazarian}, A. (2021{\natexlab{b}}).
\newblock {Anisotropies in Compressible MHD Turbulence: Probing Magnetic Fields
  and Measuring Magnetization}.
\newblock \emph{\apj} 911, 37.
\newblock \doi{10.3847/1538-4357/abea18}
\bibAnnoteFile{Huetal2021}

\bibitem[{{Hu} et~al.(2019){Hu}, {Yuen}, {Lazarian}, {Ho}, {Benjamin}, {Hill}
  et~al.}]{Huetal2019}
{Hu}, Y., {Yuen}, K.~H., {Lazarian}, V., {Ho}, K.~W., {Benjamin}, R.~A.,
  {Hill}, A.~S., et~al. (2019).
\newblock {Magnetic field morphology in interstellar clouds with the velocity
  gradient technique}.
\newblock \emph{Nature Astronomy} 3, 776--782.
\newblock \doi{10.1038/s41550-019-0769-0}
\bibAnnoteFile{Huetal2019}

\bibitem[{{Hwang} et~al.(2021){Hwang}, {Kim}, {Pattle}, {Kwon}, {Sadavoy},
  {Koch} et~al.}]{Hwangetal2021}
{Hwang}, J., {Kim}, J., {Pattle}, K., {Kwon}, W., {Sadavoy}, S., {Koch}, P.~M.,
  et~al. (2021).
\newblock {The JCMT BISTRO Survey: The Distribution of Magnetic Field Strengths
  toward the OMC-1 Region}.
\newblock \emph{\apj} 913, 85.
\newblock \doi{10.3847/1538-4357/abf3c4}
\bibAnnoteFile{Hwangetal2021}

\bibitem[{{Inoue} and {Fukui}(2013)}]{InoueFukui2013}
{Inoue}, T. and {Fukui}, Y. (2013).
\newblock {Formation of Massive Molecular Cloud Cores by Cloud-Cloud
  Collision}.
\newblock \emph{\apjl} 774, L31.
\newblock \doi{10.1088/2041-8205/774/2/L31}
\bibAnnoteFile{InoueFukui2013}

\bibitem[{{Inoue} et~al.(2018){Inoue}, {Hennebelle}, {Fukui}, {Matsumoto},
  {Iwasaki}, and {Inutsuka}}]{Inoueetal2018}
{Inoue}, T., {Hennebelle}, P., {Fukui}, Y., {Matsumoto}, T., {Iwasaki}, K., and
  {Inutsuka}, S.-i. (2018).
\newblock {The formation of massive molecular filaments and massive stars
  triggered by a magnetohydrodynamic shock wave}.
\newblock \emph{\pasj} 70, S53.
\newblock \doi{10.1093/pasj/psx089}
\bibAnnoteFile{Inoueetal2018}

\bibitem[{{Inutsuka} et~al.(2015){Inutsuka}, {Inoue}, {Iwasaki}, and
  {Hosokawa}}]{Inutsukaetal2015}
{Inutsuka}, S.-i., {Inoue}, T., {Iwasaki}, K., and {Hosokawa}, T. (2015).
\newblock {The formation and destruction of molecular clouds and galactic star
  formation. An origin for the cloud mass function and star formation
  efficiency}.
\newblock \emph{\aap} 580, A49.
\newblock \doi{10.1051/0004-6361/201425584}
\bibAnnoteFile{Inutsukaetal2015}

\bibitem[{{Inutsuka} et~al.(2016){Inutsuka}, {Inoue}, {Iwasaki}, {Hosokawa},
  and {Kobayashi}}]{Inutsuka2016IAU}
{Inutsuka}, S.-I., {Inoue}, T., {Iwasaki}, K., {Hosokawa}, T., and {Kobayashi},
  M. I.~N. (2016).
\newblock {The Formation and Destruction of Molecular Clouds and Galactic Star
  Formation}.
\newblock In \emph{From Interstellar Clouds to Star-Forming Galaxies: Universal
  Processes?}, eds. P.~{Jablonka}, P.~{Andr{\'e}}, and F.~{van der Tak}. vol.
  315, 61--68.
\newblock \doi{10.1017/S1743921316007262}
\bibAnnoteFile{Inutsuka2016IAU}

\bibitem[{{Iwasaki} et~al.(2019){Iwasaki}, {Tomida}, {Inoue}, and
  {Inutsuka}}]{Iwasakietal2019}
{Iwasaki}, K., {Tomida}, K., {Inoue}, T., and {Inutsuka}, S.-i. (2019).
\newblock {The Early Stage of Molecular Cloud Formation by Compression of
  Two-phase Atomic Gases}.
\newblock \emph{\apj} 873, 6.
\newblock \doi{10.3847/1538-4357/ab02ff}
\bibAnnoteFile{Iwasakietal2019}

\bibitem[{{Jaffe}(2019)}]{Jaffe2019}
{Jaffe}, T.~R. (2019).
\newblock {Practical Modeling of Large-Scale Galactic Magnetic Fields: Status
  and Prospects}.
\newblock \emph{Galaxies} 7, 52.
\newblock \doi{10.3390/galaxies7020052}
\bibAnnoteFile{Jaffe2019}

\bibitem[{{Jansson} and {Farrar}(2012{\natexlab{a}})}]{JanssonFarrar2012}
{Jansson}, R. and {Farrar}, G.~R. (2012{\natexlab{a}}).
\newblock {A New Model of the Galactic Magnetic Field}.
\newblock \emph{\apj} 757, 14.
\newblock \doi{10.1088/0004-637X/757/1/14}
\bibAnnoteFile{JanssonFarrar2012}

\bibitem[{{Jansson} and {Farrar}(2012{\natexlab{b}})}]{JanssonFarrar2012b}
{Jansson}, R. and {Farrar}, G.~R. (2012{\natexlab{b}}).
\newblock {The Galactic Magnetic Field}.
\newblock \emph{\apjl} 761, L11.
\newblock \doi{10.1088/2041-8205/761/1/L11}
\bibAnnoteFile{JanssonFarrar2012b}

\bibitem[{{Kainulainen} et~al.(2009){Kainulainen}, {Beuther}, {Henning}, and
  {Plume}}]{Kainulainenetal2009}
{Kainulainen}, J., {Beuther}, H., {Henning}, T., and {Plume}, R. (2009).
\newblock {Probing the evolution of molecular cloud structure. From quiescence
  to birth}.
\newblock \emph{\aap} 508, L35--L38.
\newblock \doi{10.1051/0004-6361/200913605}
\bibAnnoteFile{Kainulainenetal2009}

\bibitem[{{Kandori} et~al.(2020{\natexlab{a}}){Kandori}, {Tamura}, {Saito},
  {Tomisaka}, {Matsumoto}, {Kusakabe} et~al.}]{Kandorietal2020_3Da}
{Kandori}, R., {Tamura}, M., {Saito}, M., {Tomisaka}, K., {Matsumoto}, T.,
  {Kusakabe}, N., et~al. (2020{\natexlab{a}}).
\newblock {Distortion of magnetic fields in Barnard 68}.
\newblock \emph{\pasj} 72, 8.
\newblock \doi{10.1093/pasj/psz127}
\bibAnnoteFile{Kandorietal2020_3Da}

\bibitem[{{Kandori} et~al.(2020{\natexlab{b}}){Kandori}, {Tamura}, {Saito},
  {Tomisaka}, {Matsumoto}, {Tazaki} et~al.}]{Kandorietal2020_3Db}
{Kandori}, R., {Tamura}, M., {Saito}, M., {Tomisaka}, K., {Matsumoto}, T.,
  {Tazaki}, R., et~al. (2020{\natexlab{b}}).
\newblock {Distortion of Magnetic Fields in the Dense Core CB81 (L1774, Pipe
  42) in the Pipe Nebula}.
\newblock \emph{\apj} 890, 14.
\newblock \doi{10.3847/1538-4357/ab67c5}
\bibAnnoteFile{Kandorietal2020_3Db}

\bibitem[{{Kandori} et~al.(2017){Kandori}, {Tamura}, {Tomisaka}, {Nakajima},
  {Kusakabe}, {Kwon} et~al.}]{Kandorietal2017}
{Kandori}, R., {Tamura}, M., {Tomisaka}, K., {Nakajima}, Y., {Kusakabe}, N.,
  {Kwon}, J., et~al. (2017).
\newblock {Distortion of Magnetic Fields in a Starless Core II: 3D Magnetic
  Field Structure of FeSt 1-457}.
\newblock \emph{\apj} 848, 110.
\newblock \doi{10.3847/1538-4357/aa8d18}
\bibAnnoteFile{Kandorietal2017}

\bibitem[{{Kandori} et~al.(2020{\natexlab{c}}){Kandori}, {Tomisaka}, {Saito},
  {Tamura}, {Matsumoto}, {Tazaki} et~al.}]{Kandorietal2020_3Dc}
{Kandori}, R., {Tomisaka}, K., {Saito}, M., {Tamura}, M., {Matsumoto}, T.,
  {Tazaki}, R., et~al. (2020{\natexlab{c}}).
\newblock {Distortion of Magnetic Fields in a Starless Core. VI. Application of
  Flux Freezing Model and Core Formation of FeSt 1-457}.
\newblock \emph{\apj} 888, 120.
\newblock \doi{10.3847/1538-4357/ab6081}
\bibAnnoteFile{Kandorietal2020_3Dc}

\bibitem[{Kim and Ostriker(2015)}]{KimOstriker2015}
Kim, C.-G. and Ostriker, E.~C. (2015).
\newblock {MOMENTUM} {INJECTION} {BY} {SUPERNOVAE} {IN} {THE} {INTERSTELLAR}
  {MEDIUM}.
\newblock \emph{The Astrophysical Journal} 802, 99.
\newblock \doi{10.1088/0004-637x/802/2/99}
\bibAnnoteFile{KimOstriker2015}

\bibitem[{{King} et~al.(2019){King}, {Chen}, {Fissel}, and {Li}}]{Kingetal2019}
{King}, P.~K., {Chen}, C.-Y., {Fissel}, L.~M., and {Li}, Z.-Y. (2019).
\newblock {Effects of grain alignment efficiency on synthetic dust polarization
  observations of molecular clouds}.
\newblock \emph{\mnras} 490, 2760--2778.
\newblock \doi{10.1093/mnras/stz2628}
\bibAnnoteFile{Kingetal2019}

\bibitem[{{King} et~al.(2018){King}, {Fissel}, {Chen}, and {Li}}]{Kingetal2018}
{King}, P.~K., {Fissel}, L.~M., {Chen}, C.-Y., and {Li}, Z.-Y. (2018).
\newblock {Modelling dust polarization observations of molecular clouds through
  MHD simulations}.
\newblock \emph{\mnras} 474, 5122--5142.
\newblock \doi{10.1093/mnras/stx3096}
\bibAnnoteFile{Kingetal2018}

\bibitem[{{Kounkel} and {Covey}(2019)}]{KounkelCovey2019}
{Kounkel}, M. and {Covey}, K. (2019).
\newblock {Untangling the Galaxy. I. Local Structure and Star Formation History
  of the Milky Way}.
\newblock \emph{\aj} 158, 122.
\newblock \doi{10.3847/1538-3881/ab339a}
\bibAnnoteFile{KounkelCovey2019}

\bibitem[{{Kounkel} et~al.(2022){Kounkel}, {Deng}, and
  {Stassun}}]{Kounkeletal2022}
{Kounkel}, M., {Deng}, T., and {Stassun}, K.~G. (2022).
\newblock {Dynamical star forming history of Per OB2}.
\newblock \emph{arXiv e-prints} , arXiv:2206.04703
\bibAnnoteFile{Kounkeletal2022}

\bibitem[{{Krumholz} and {Federrath}(2019)}]{KrumholzFederrath2019}
{Krumholz}, M.~R. and {Federrath}, C. (2019).
\newblock {The Role of Magnetic Fields in Setting the Star Formation Rate and
  the Initial Mass Function}.
\newblock \emph{Frontiers in Astronomy and Space Sciences} 6, 7.
\newblock \doi{10.3389/fspas.2019.00007}
\bibAnnoteFile{KrumholzFederrath2019}

\bibitem[{{Kwon} et~al.(2022){Kwon}, {Pattle}, {Sadavoy}, {Hull}, {Johnstone},
  {Ward-Thompson} et~al.}]{Kwonetal2022}
{Kwon}, W., {Pattle}, K., {Sadavoy}, S., {Hull}, C. L.~H., {Johnstone}, D.,
  {Ward-Thompson}, D., et~al. (2022).
\newblock {B-fields in Star-forming Region Observations (BISTRO): Magnetic
  Fields in the Filamentary Structures of Serpens Main}.
\newblock \emph{\apj} 926, 163.
\newblock \doi{10.3847/1538-4357/ac4bbe}
\bibAnnoteFile{Kwonetal2022}

\bibitem[{{Lazarian}(2007)}]{Lazarian2007}
{Lazarian}, A. (2007).
\newblock {Tracing magnetic fields with aligned grains}.
\newblock \emph{\jqsrt} 106, 225--256.
\newblock \doi{10.1016/j.jqsrt.2007.01.038}
\bibAnnoteFile{Lazarian2007}

\bibitem[{{Lazarian} and {Hoang}(2007)}]{LazarianHoang2007}
{Lazarian}, A. and {Hoang}, T. (2007).
\newblock {Radiative torques: analytical model and basic properties}.
\newblock \emph{\mnras} 378, 910--946.
\newblock \doi{10.1111/j.1365-2966.2007.11817.x}
\bibAnnoteFile{LazarianHoang2007}

\bibitem[{{Lazarian} and {Yuen}(2018)}]{LazarianYuen2018}
{Lazarian}, A. and {Yuen}, K.~H. (2018).
\newblock {Gradients of Synchrotron Polarization: Tracing 3D Distribution of
  Magnetic Fields}.
\newblock \emph{\apj} 865, 59.
\newblock \doi{10.3847/1538-4357/aad3ca}
\bibAnnoteFile{LazarianYuen2018}

\bibitem[{{Li} and {Houde}(2008)}]{LiHoude2008}
{Li}, H.-b. and {Houde}, M. (2008).
\newblock {Probing the Turbulence Dissipation Range and Magnetic Field
  Strengths in Molecular Clouds}.
\newblock \emph{\apj} 677, 1151--1156.
\newblock \doi{10.1086/529581}
\bibAnnoteFile{LiHoude2008}

\bibitem[{{Lombardi} et~al.(2014){Lombardi}, {Bouy}, {Alves}, and
  {Lada}}]{Lombardi2014}
{Lombardi}, M., {Bouy}, H., {Alves}, J., and {Lada}, C.~J. (2014).
\newblock {Herschel-Planck dust optical-depth and column-density maps. I.
  Method description and results for Orion}.
\newblock \emph{\aap} 566, A45.
\newblock \doi{10.1051/0004-6361/201323293}
\bibAnnoteFile{Lombardi2014}

\bibitem[{{Luri} et~al.(2018){Luri}, {Brown}, {Sarro}, {Arenou},
  {Bailer-Jones}, {Castro-Ginard} et~al.}]{Lurietal2018}
{Luri}, X., {Brown}, A.~G.~A., {Sarro}, L.~M., {Arenou}, F., {Bailer-Jones},
  C.~A.~L., {Castro-Ginard}, A., et~al. (2018).
\newblock {Gaia Data Release 2. Using Gaia parallaxes}.
\newblock \emph{\aap} 616, A9.
\newblock \doi{10.1051/0004-6361/201832964}
\bibAnnoteFile{Lurietal2018}

\bibitem[{{Mouschovias} and {Tassis}(2010)}]{MouschoviasTassis2010}
{Mouschovias}, T.~C. and {Tassis}, K. (2010).
\newblock {Self-consistent analysis of OH-Zeeman observations: too much noise
  about noise}.
\newblock \emph{\mnras} 409, 801--807.
\newblock \doi{10.1111/j.1365-2966.2010.17345.x}
\bibAnnoteFile{MouschoviasTassis2010}

\bibitem[{{Ngoc} et~al.(2021){Ngoc}, {Diep}, {Parsons}, {Pattle}, {Hoang},
  {Ward-Thompson} et~al.}]{Ngocetal2021}
{Ngoc}, N.~B., {Diep}, P.~N., {Parsons}, H., {Pattle}, K., {Hoang}, T.,
  {Ward-Thompson}, D., et~al. (2021).
\newblock {Observations of Magnetic Fields Surrounding LkH{\ensuremath{\alpha}}
  101 Taken by the BISTRO Survey with JCMT-POL-2}.
\newblock \emph{\apj} 908, 10.
\newblock \doi{10.3847/1538-4357/abd0fc}
\bibAnnoteFile{Ngocetal2021}

\bibitem[{{Noutsos}(2012)}]{Noutsos2012}
{Noutsos}, A. (2012).
\newblock {The Magnetic Field of the Milky Way from Faraday Rotation of Pulsars
  and Extragalactic Sources}.
\newblock \emph{\ssr} 166, 307--324.
\newblock \doi{10.1007/s11214-011-9860-2}
\bibAnnoteFile{Noutsos2012}

\bibitem[{{Ostriker} et~al.(2001){Ostriker}, {Stone}, and
  {Gammie}}]{Ostrikeretal2001}
{Ostriker}, E.~C., {Stone}, J.~M., and {Gammie}, C.~F. (2001).
\newblock {Density, Velocity, and Magnetic Field Structure in Turbulent
  Molecular Cloud Models}.
\newblock \emph{\apj} 546, 980--1005.
\newblock \doi{10.1086/318290}
\bibAnnoteFile{Ostrikeretal2001}

\bibitem[{{Panopoulou} et~al.(2019){Panopoulou}, {Tassis}, {Skalidis},
  {Blinov}, {Liodakis}, {Pavlidou} et~al.}]{Panopoulouetal2019}
{Panopoulou}, G.~V., {Tassis}, K., {Skalidis}, R., {Blinov}, D., {Liodakis},
  I., {Pavlidou}, V., et~al. (2019).
\newblock {Demonstration of Magnetic Field Tomography with Starlight
  Polarization toward a Diffuse Sightline of the ISM}.
\newblock \emph{\apj} 872, 56.
\newblock \doi{10.3847/1538-4357/aafdb2}
\bibAnnoteFile{Panopoulouetal2019}

\bibitem[{{Pattle} and {Fissel}(2019)}]{PattleFissel2019}
{Pattle}, K. and {Fissel}, L. (2019).
\newblock {Submillimeter and Far-infrared Polarimetric Observations of Magnetic
  Fields in Star-Forming Regions}.
\newblock \emph{Frontiers in Astronomy and Space Sciences} 6, 15.
\newblock \doi{10.3389/fspas.2019.00015}
\bibAnnoteFile{PattleFissel2019}

\bibitem[{{Pattle} et~al.(2022){Pattle}, {Fissel}, {Tahani}, {Liu}, and
  {Ntormousi}}]{Pattleetal2022PP7}
{Pattle}, K., {Fissel}, L., {Tahani}, M., {Liu}, T., and {Ntormousi}, E.
  (2022).
\newblock {Magnetic fields in star formation: from clouds to cores}.
\newblock \emph{arXiv e-prints} , arXiv:2203.11179
\bibAnnoteFile{Pattleetal2022PP7}

\bibitem[{{Pattle} et~al.(2021){Pattle}, {Lai}, {Wright}, {Coud{\'e}},
  {Plambeck}, {Hoang} et~al.}]{Pattleetal2021}
{Pattle}, K., {Lai}, S.-P., {Wright}, M., {Coud{\'e}}, S., {Plambeck}, R.,
  {Hoang}, T., et~al. (2021).
\newblock {OMC-1 dust polarization in ALMA Band 7: diagnosing grain alignment
  mechanisms in the vicinity of Orion Source I}.
\newblock \emph{\mnras} 503, 3414--3433.
\newblock \doi{10.1093/mnras/stab608}
\bibAnnoteFile{Pattleetal2021}

\bibitem[{{Pereyra} and {Magalh{\~a}es}(2007)}]{PereyraMagalhaes2007}
{Pereyra}, A. and {Magalh{\~a}es}, A.~M. (2007).
\newblock {Polarimetry toward the IRAS Vela Shell. II. Extinction and Magnetic
  Fields}.
\newblock \emph{\apj} 662, 1014--1023.
\newblock \doi{10.1086/517906}
\bibAnnoteFile{PereyraMagalhaes2007}

\bibitem[{{Pillai} et~al.(2016){Pillai}, {Kauffmann}, {Wiesemeyer}, and
  {Menten}}]{Pillaietal2016}
{Pillai}, T., {Kauffmann}, J., {Wiesemeyer}, H., and {Menten}, K.~M. (2016).
\newblock {CN Zeeman and dust polarization in a high-mass cold clump}.
\newblock \emph{\aap} 591, A19.
\newblock \doi{10.1051/0004-6361/201527803}
\bibAnnoteFile{Pillaietal2016}

\bibitem[{{Pineda} et~al.(2021){Pineda}, {Schmiedeke}, {Caselli}, {Stahler},
  {Frayer}, {Church} et~al.}]{Pinedaetal2021}
{Pineda}, J.~E., {Schmiedeke}, A., {Caselli}, P., {Stahler}, S.~W., {Frayer},
  D.~T., {Church}, S.~E., et~al. (2021).
\newblock {Neutral versus Ion Line Widths in Barnard 5: Evidence for
  Penetration by Magnetohydrodynamic Waves}.
\newblock \emph{\apj} 912, 7.
\newblock \doi{10.3847/1538-4357/abebdd}
\bibAnnoteFile{Pinedaetal2021}

\bibitem[{{Planck Collaboration} et~al.(2016){Planck Collaboration}, {Ade},
  {Aghanim}, {Alves}, {Arnaud}, {Arzoumanian} et~al.}]{PlanckXXXV}
{Planck Collaboration}, {Ade}, P.~A.~R., {Aghanim}, N., {Alves}, M.~I.~R.,
  {Arnaud}, M., {Arzoumanian}, D., et~al. (2016).
\newblock {Planck intermediate results. XXXV. Probing the role of the magnetic
  field in the formation of structure in molecular clouds}.
\newblock \emph{\aap} 586, A138.
\newblock \doi{10.1051/0004-6361/201525896}
\bibAnnoteFile{PlanckXXXV}

\bibitem[{{Poggio} et~al.(2020){Poggio}, {Drimmel}, {Andrae}, {Bailer-Jones},
  {Fouesneau}, {Lattanzi} et~al.}]{Poggioetal2020}
{Poggio}, E., {Drimmel}, R., {Andrae}, R., {Bailer-Jones}, C.~A.~L.,
  {Fouesneau}, M., {Lattanzi}, M.~G., et~al. (2020).
\newblock {Evidence of a dynamically evolving Galactic warp}.
\newblock \emph{Nature Astronomy} 4, 590--596.
\newblock \doi{10.1038/s41550-020-1017-3}
\bibAnnoteFile{Poggioetal2020}

\bibitem[{{Reich} et~al.(2002){Reich}, {F{\"u}rst}, {Reich}, {Wielebinski}, and
  {Wolleben}}]{Reichetal2002}
{Reich}, W., {F{\"u}rst}, E., {Reich}, P., {Wielebinski}, R., and {Wolleben},
  M. (2002).
\newblock {Polarization surveys of the galaxy}.
\newblock In \emph{Astrophysical Polarized Backgrounds}, eds. S.~{Cecchini},
  S.~{Cortiglioni}, R.~{Sault}, and C.~{Sbarra}. vol. 609 of \emph{American
  Institute of Physics Conference Series}, 3--8.
\newblock \doi{10.1063/1.1471815}
\bibAnnoteFile{Reichetal2002}

\bibitem[{{Rezaei Kh.} et~al.(2020){Rezaei Kh.}, {Bailer-Jones}, {Soler}, and
  {Zari}}]{Rezaeietal2020}
{Rezaei Kh.}, S., {Bailer-Jones}, C. A.~L., {Soler}, J.~D., and {Zari}, E.
  (2020).
\newblock {Detailed 3D structure of Orion A in dust with Gaia DR2}.
\newblock \emph{\aap} 643, A151.
\newblock \doi{10.1051/0004-6361/202038708}
\bibAnnoteFile{Rezaeietal2020}

\bibitem[{{Rezaei Kh.} and {Kainulainen}(2022)}]{RezaeiKainulainen2022}
{Rezaei Kh.}, S. and {Kainulainen}, J. (2022).
\newblock {Three-dimensional Shape Explains Star Formation Mystery of
  California and Orion A}.
\newblock \emph{\apjl} 930, L22.
\newblock \doi{10.3847/2041-8213/ac67db}
\bibAnnoteFile{RezaeiKainulainen2022}

\bibitem[{{Robishaw} et~al.(2015){Robishaw}, {Green}, {Surcis}, {Vlemmings},
  {Richards}, {Etoka} et~al.}]{Robishawetal2015}
{Robishaw}, T., {Green}, J., {Surcis}, G., {Vlemmings}, W.~H.~T., {Richards},
  A.~M.~S., {Etoka}, S., et~al. (2015).
\newblock {Measuring Magnetic Fields Near and Far with the SKA via the Zeeman
  Effect}.
\newblock In \emph{Advancing Astrophysics with the Square Kilometre Array
  (AASKA14)}. 110
\bibAnnoteFile{Robishawetal2015}

\bibitem[{{Ruiz-Lara} et~al.(2020){Ruiz-Lara}, {Gallart}, {Bernard}, and
  {Cassisi}}]{RuizLaraetal2020}
{Ruiz-Lara}, T., {Gallart}, C., {Bernard}, E.~J., and {Cassisi}, S. (2020).
\newblock {The recurrent impact of the Sagittarius dwarf on the star formation
  history of the Milky Way}.
\newblock \emph{Nature Astronomy} 4, 965--973.
\newblock \doi{10.1038/s41550-020-1097-0}
\bibAnnoteFile{RuizLaraetal2020}

\bibitem[{{Seifried} and {Walch}(2015)}]{SeifriedWalch2015}
{Seifried}, D. and {Walch}, S. (2015).
\newblock {The impact of turbulence and magnetic field orientation on
  star-forming filaments}.
\newblock \emph{\mnras} 452, 2410--2422.
\newblock \doi{10.1093/mnras/stv1458}
\bibAnnoteFile{SeifriedWalch2015}

\bibitem[{{Skalidis} et~al.(2021{\natexlab{a}}){Skalidis}, {Sternberg},
  {Beattie}, {Pavlidou}, and {Tassis}}]{Skalidisetal2021a}
{Skalidis}, R., {Sternberg}, J., {Beattie}, J.~R., {Pavlidou}, V., and
  {Tassis}, K. (2021{\natexlab{a}}).
\newblock {Why take the square root? An assessment of interstellar magnetic
  field strength estimation methods}.
\newblock \emph{\aap} 656, A118.
\newblock \doi{10.1051/0004-6361/202142045}
\bibAnnoteFile{Skalidisetal2021a}

\bibitem[{{Skalidis} and {Tassis}(2021)}]{SkalidisTassis2021}
{Skalidis}, R. and {Tassis}, K. (2021).
\newblock {High-accuracy estimation of magnetic field strength in the
  interstellar medium from dust polarization}.
\newblock \emph{\aap} 647, A186.
\newblock \doi{10.1051/0004-6361/202039779}
\bibAnnoteFile{SkalidisTassis2021}

\bibitem[{{Skalidis} et~al.(2021{\natexlab{b}}){Skalidis}, {Tassis},
  {Panopoulou}, {Pineda}, {Gong}, {Mandarakas} et~al.}]{Skalidisetal2021}
{Skalidis}, R., {Tassis}, K., {Panopoulou}, G.~V., {Pineda}, J.~L., {Gong}, Y.,
  {Mandarakas}, N., et~al. (2021{\natexlab{b}}).
\newblock {HI-H$_2$ transition: exploring the role of the magnetic field}.
\newblock \emph{arXiv e-prints} , arXiv:2110.11878
\bibAnnoteFile{Skalidisetal2021}

\bibitem[{{Soler}(2019)}]{Soler2019}
{Soler}, J.~D. (2019).
\newblock {Using Herschel and Planck observations to delineate the role of
  magnetic fields in molecular cloud structure}.
\newblock \emph{\aap} 629, A96.
\newblock \doi{10.1051/0004-6361/201935779}
\bibAnnoteFile{Soler2019}

\bibitem[{{Soler} et~al.(2017){Soler}, {Ade}, {Angil{\`e}}, {Ashton}, {Benton},
  {Devlin} et~al.}]{Soleretal2017}
{Soler}, J.~D., {Ade}, P.~A.~R., {Angil{\`e}}, F.~E., {Ashton}, P., {Benton},
  S.~J., {Devlin}, M.~J., et~al. (2017).
\newblock {The relation between the column density structures and the magnetic
  field orientation in the Vela C molecular complex}.
\newblock \emph{\aap} 603, A64.
\newblock \doi{10.1051/0004-6361/201730608}
\bibAnnoteFile{Soleretal2017}

\bibitem[{{Soler} et~al.(2018){Soler}, {Bracco}, and {Pon}}]{Soleretal2018}
{Soler}, J.~D., {Bracco}, A., and {Pon}, A. (2018).
\newblock {The magnetic environment of the Orion-Eridanus superbubble as
  revealed by Planck}.
\newblock \emph{\aap} 609, L3.
\newblock \doi{10.1051/0004-6361/201732203}
\bibAnnoteFile{Soleretal2018}

\bibitem[{{Soler} and {Hennebelle}(2017)}]{SolerHennebelle2017}
{Soler}, J.~D. and {Hennebelle}, P. (2017).
\newblock {What are we learning from the relative orientation between density
  structures and the magnetic field in molecular clouds?}
\newblock \emph{\aap} 607, A2.
\newblock \doi{10.1051/0004-6361/201731049}
\bibAnnoteFile{SolerHennebelle2017}

\bibitem[{{Soubiran} et~al.(2018){Soubiran}, {Jasniewicz}, {Chemin}, {Zurbach},
  {Brouillet}, {Panuzzo} et~al.}]{Soubiranetal2018}
{Soubiran}, C., {Jasniewicz}, G., {Chemin}, L., {Zurbach}, C., {Brouillet}, N.,
  {Panuzzo}, P., et~al. (2018).
\newblock {Gaia Data Release 2. The catalogue of radial velocity standard
  stars}.
\newblock \emph{\aap} 616, A7.
\newblock \doi{10.1051/0004-6361/201832795}
\bibAnnoteFile{Soubiranetal2018}

\bibitem[{{Stephens} et~al.(2011){Stephens}, {Looney}, {Dowell},
  {Vaillancourt}, and {Tassis}}]{Stephensetal2011}
{Stephens}, I.~W., {Looney}, L.~W., {Dowell}, C.~D., {Vaillancourt}, J.~E., and
  {Tassis}, K. (2011).
\newblock {The Galactic Magnetic Field's Effect in Star-forming Regions}.
\newblock \emph{\apj} 728, 99.
\newblock \doi{10.1088/0004-637X/728/2/99}
\bibAnnoteFile{Stephensetal2011}

\bibitem[{{Stone} et~al.(2008){Stone}, {Gardiner}, {Teuben}, {Hawley}, and
  {Simon}}]{Stoneetal2008}
{Stone}, J.~M., {Gardiner}, T.~A., {Teuben}, P., {Hawley}, J.~F., and {Simon},
  J.~B. (2008).
\newblock {Athena: A New Code for Astrophysical MHD}.
\newblock \emph{\apjs} 178, 137--177.
\newblock \doi{10.1086/588755}
\bibAnnoteFile{Stoneetal2008}

\bibitem[{{Sullivan} et~al.(2021){Sullivan}, {Fissel}, {King}, {Chen}, {Li},
  and {Soler}}]{Sullivan2021}
{Sullivan}, C.~H., {Fissel}, L.~M., {King}, P.~K., {Chen}, C.~Y., {Li}, Z.~Y.,
  and {Soler}, J.~D. (2021).
\newblock {Characterizing the magnetic fields of nearby molecular clouds using
  submillimeter polarization observations}.
\newblock \emph{\mnras} 503, 5006--5024.
\newblock \doi{10.1093/mnras/stab596}
\bibAnnoteFile{Sullivan2021}

\bibitem[{{Tahani} et~al.(2022{\natexlab{a}}){Tahani}, {Glover}, {Lupypciw},
  {West}, {Kothes}, {Plume} et~al.}]{Tahanietal2022O}
{Tahani}, M., {Glover}, J., {Lupypciw}, W., {West}, J.~L., {Kothes}, R.,
  {Plume}, R., et~al. (2022{\natexlab{a}}).
\newblock {Orion A's complete 3D magnetic field morphology}.
\newblock \emph{\aap} 660, L7.
\newblock \doi{10.1051/0004-6361/202243322}
\bibAnnoteFile{Tahanietal2022O}

\bibitem[{{Tahani} et~al.(2022{\natexlab{b}}){Tahani}, {Lupypciw}, {Glover},
  {Plume}, {West}, {Kothes} et~al.}]{Tahanietal2022P}
{Tahani}, M., {Lupypciw}, W., {Glover}, J., {Plume}, R., {West}, J.~L.,
  {Kothes}, R., et~al. (2022{\natexlab{b}}).
\newblock {3D magnetic field morphology of the Perseus molecular cloud}.
\newblock \emph{arXiv e-prints} , arXiv:2201.04718
\bibAnnoteFile{Tahanietal2022P}

\bibitem[{{Tahani} et~al.(2018){Tahani}, {Plume}, {Brown}, and
  {Kainulainen}}]{Tahanietal2018}
{Tahani}, M., {Plume}, R., {Brown}, J.~C., and {Kainulainen}, J. (2018).
\newblock {Helical magnetic fields in molecular clouds?. A new method to
  determine the line-of-sight magnetic field structure in molecular clouds}.
\newblock \emph{\aap} 614, A100.
\newblock \doi{10.1051/0004-6361/201732219}
\bibAnnoteFile{Tahanietal2018}

\bibitem[{{Tahani} et~al.(2019){Tahani}, {Plume}, {Brown}, {Soler}, and
  {Kainulainen}}]{Tahanietal2019}
{Tahani}, M., {Plume}, R., {Brown}, J.~C., {Soler}, J.~D., and {Kainulainen},
  J. (2019).
\newblock {Could bow-shaped magnetic morphologies surround filamentary
  molecular clouds?. The 3D magnetic field structure of Orion-A}.
\newblock \emph{\aap} 632, A68.
\newblock \doi{10.1051/0004-6361/201936280}
\bibAnnoteFile{Tahanietal2019}

\bibitem[{{Tang} et~al.(2018){Tang}, {Li}, and {Lee}}]{Tangetal2018}
{Tang}, K.~S., {Li}, H.-B., and {Lee}, W.-K. (2018).
\newblock {Probing the Turbulence Dissipation Range and Magnetic Field
  Strengths in Molecular Clouds. II. Directly Probing the Ion-neutral
  Decoupling Scale}.
\newblock \emph{\apj} 862, 42.
\newblock \doi{10.3847/1538-4357/aacb82}
\bibAnnoteFile{Tangetal2018}

\bibitem[{{Tassis} et~al.(2018){Tassis}, {Ramaprakash}, {Readhead}, {Potter},
  {Wehus}, {Panopoulou} et~al.}]{Tassisetal2018}
{Tassis}, K., {Ramaprakash}, A.~N., {Readhead}, A. C.~S., {Potter}, S.~B.,
  {Wehus}, I.~K., {Panopoulou}, G.~V., et~al. (2018).
\newblock {PASIPHAE: A high-Galactic-latitude, high-accuracy optopolarimetric
  survey}.
\newblock \emph{arXiv e-prints} , arXiv:1810.05652
\bibAnnoteFile{Tassisetal2018}

\bibitem[{{Taylor} et~al.(2009){Taylor}, {Stil}, and
  {Sunstrum}}]{Tayloretal2009}
{Taylor}, A.~R., {Stil}, J.~M., and {Sunstrum}, C. (2009).
\newblock {A Rotation Measure Image of the Sky}.
\newblock \emph{\apj} 702, 1230--1236.
\newblock \doi{10.1088/0004-637X/702/2/1230}
\bibAnnoteFile{Tayloretal2009}

\bibitem[{{Tritsis} and {Tassis}(2018)}]{TritsisTassis2018}
{Tritsis}, A. and {Tassis}, K. (2018).
\newblock {Magnetic seismology of interstellar gas clouds: Unveiling a hidden
  dimension}.
\newblock \emph{Science} 360, 635--638.
\newblock \doi{10.1126/science.aao1185}
\bibAnnoteFile{TritsisTassis2018}

\bibitem[{{Van Eck} et~al.(2017){Van Eck}, {Haverkorn}, {Alves}, {Beck}, {de
  Bruyn}, {En{\ss}lin} et~al.}]{VanEcketal2017}
{Van Eck}, C.~L., {Haverkorn}, M., {Alves}, M.~I.~R., {Beck}, R., {de Bruyn},
  A.~G., {En{\ss}lin}, T., et~al. (2017).
\newblock {Faraday tomography of the local interstellar medium with LOFAR:
  Galactic foregrounds towards IC 342}.
\newblock \emph{\aap} 597, A98.
\newblock \doi{10.1051/0004-6361/201629707}
\bibAnnoteFile{VanEcketal2017}

\bibitem[{{Voshchinnikov}(2012)}]{Voshchinnikov2012}
{Voshchinnikov}, N.~V. (2012).
\newblock {Interstellar extinction and interstellar polarization: Old and new
  models}.
\newblock \emph{\jqsrt} 113, 2334--2350.
\newblock \doi{10.1016/j.jqsrt.2012.06.013}
\bibAnnoteFile{Voshchinnikov2012}

\bibitem[{{Wolleben} and {Reich}(2004)}]{WollebenReich2004}
{Wolleben}, M. and {Reich}, W. (2004).
\newblock {Modelling Faraday Screens in the Interstellar Medium}.
\newblock In \emph{The Magnetized Interstellar Medium}, eds. B.~{Uyaniker},
  W.~{Reich}, and R.~{Wielebinski}. 99--104
\bibAnnoteFile{WollebenReich2004}

\bibitem[{{Wyrzykowski} et~al.(2020){Wyrzykowski}, {Mr{\'o}z}, {Rybicki},
  {Gromadzki}, {Ko{\l}aczkowski}, {Zieli{\'n}ski} et~al.}]{Wyrzykowskietal2020}
{Wyrzykowski}, {\L}., {Mr{\'o}z}, P., {Rybicki}, K.~A., {Gromadzki}, M.,
  {Ko{\l}aczkowski}, Z., {Zieli{\'n}ski}, M., et~al. (2020).
\newblock {Full orbital solution for the binary system in the northern Galactic
  disc microlensing event Gaia16aye}.
\newblock \emph{\aap} 633, A98.
\newblock \doi{10.1051/0004-6361/201935097}
\bibAnnoteFile{Wyrzykowskietal2020}

\bibitem[{{Yan} et~al.(2019){Yan}, {Gry}, {Boulanger}, and
  {Leone}}]{Yanetal2019}
{Yan}, H., {Gry}, C., {Boulanger}, F., and {Leone}, F. (2019).
\newblock {Precision measurement of magnetic field from near to far, from fine
  to large scales in ISM}.
\newblock \emph{\baas} 51, 217
\bibAnnoteFile{Yanetal2019}

\bibitem[{{Yan} and {Lazarian}(2005)}]{YanLazarian2005}
{Yan}, H. and {Lazarian}, A. (2005).
\newblock {Optical Polarization from Aligned Atoms as a New Diagnostic of
  Astrophysical Magnetic fields}.
\newblock In \emph{Astronomical Polarimetry: Current Status and Future
  Directions}, eds. A.~{Adamson}, C.~{Aspin}, C.~{Davis}, and T.~{Fujiyoshi}.
  vol. 343 of \emph{Astronomical Society of the Pacific Conference Series}, 346
\bibAnnoteFile{YanLazarian2005}

\bibitem[{{Yan} and {Lazarian}(2006)}]{YanLazarian2006}
{Yan}, H. and {Lazarian}, A. (2006).
\newblock {Polarization of Absorption Lines as a Diagnostics of Circumstellar,
  Interstellar, and Intergalactic Magnetic Fields: Fine-Structure Atoms}.
\newblock \emph{\apj} 653, 1292--1313.
\newblock \doi{10.1086/508704}
\bibAnnoteFile{YanLazarian2006}

\bibitem[{{Yan} and {Lazarian}(2007)}]{YanLazarian2007}
{Yan}, H. and {Lazarian}, A. (2007).
\newblock {Polarization from Aligned Atoms as a Diagnostic of Circumstellar,
  Active Galactic Nuclei, and Interstellar Magnetic Fields. II. Atoms with
  Hyperfine Structure}.
\newblock \emph{\apj} 657, 618--640.
\newblock \doi{10.1086/510847}
\bibAnnoteFile{YanLazarian2007}

\bibitem[{{Yan} and {Lazarian}(2012)}]{YanLazarian2012}
{Yan}, H. and {Lazarian}, A. (2012).
\newblock {Tracing magnetic fields with ground state alignment}.
\newblock \emph{\jqsrt} 113, 1409--1428.
\newblock \doi{10.1016/j.jqsrt.2012.03.027}
\bibAnnoteFile{YanLazarian2012}

\bibitem[{{Zari} et~al.(2016){Zari}, {Lombardi}, {Alves}, {Lada}, and
  {Bouy}}]{Zarietal2016}
{Zari}, E., {Lombardi}, M., {Alves}, J., {Lada}, C.~J., and {Bouy}, H. (2016).
\newblock {Herschel-Planck dust optical depth and column density maps. II.
  Perseus}.
\newblock \emph{\aap} 587, A106.
\newblock \doi{10.1051/0004-6361/201526597}
\bibAnnoteFile{Zarietal2016}

\bibitem[{{Zhang} et~al.(2014){Zhang}, {Qiu}, {Girart}, {Liu}, {Tang}, {Koch}
  et~al.}]{Zhangetal2014}
{Zhang}, Q., {Qiu}, K., {Girart}, J.~M., {Liu}, H.~B., {Tang}, Y.-W., {Koch},
  P.~M., et~al. (2014).
\newblock {Magnetic Fields and Massive Star Formation}.
\newblock \emph{\apj} 792, 116.
\newblock \doi{10.1088/0004-637X/792/2/116}
\bibAnnoteFile{Zhangetal2014}

\bibitem[{{Zucker} et~al.(2021){Zucker}, {Goodman}, {Alves}, {Bialy}, {Koch},
  {Speagle} et~al.}]{Zuckeretal2021}
{Zucker}, C., {Goodman}, A., {Alves}, J., {Bialy}, S., {Koch}, E.~W.,
  {Speagle}, J.~S., et~al. (2021).
\newblock {On the Three-dimensional Structure of Local Molecular Clouds}.
\newblock \emph{\apj} 919, 35.
\newblock \doi{10.3847/1538-4357/ac1f96}
\bibAnnoteFile{Zuckeretal2021}

\bibitem[{{Zucker} et~al.(2022){Zucker}, {Goodman}, {Alves}, {Bialy}, {Foley},
  {Speagle} et~al.}]{Zuckeretal2022}
{Zucker}, C., {Goodman}, A.~A., {Alves}, J., {Bialy}, S., {Foley}, M.,
  {Speagle}, J.~S., et~al. (2022).
\newblock {Star formation near the Sun is driven by expansion of the Local
  Bubble}.
\newblock \emph{\nat} 601, 334--337.
\newblock \doi{10.1038/s41586-021-04286-5}
\bibAnnoteFile{Zuckeretal2022}

\end{thebibliography}
\bibliographystyle{Frontiers-Harvard}

\end{document}